%% file: main.tex
\documentclass[11pt]{article}
\usepackage{graphicx} 
\usepackage{amsmath}
\usepackage{amssymb}
\usepackage{algpseudocode}
\usepackage{algorithm}
\usepackage[colorlinks=true, allcolors=blue]{hyperref}
\usepackage[style=numeric-comp,sorting=none]{biblatex}
\usepackage{authblk}
\usepackage{booktabs}
\usepackage{adjustbox}
\usepackage{mathtools}

\newcommand{\beginsupplement}{%
        \setcounter{table}{0}
        \renewcommand{\thetable}{S\arabic{table}}%
        \setcounter{figure}{0}
        \renewcommand{\thefigure}{S\arabic{figure}}%
     }

\title{Bayesian Inference for Epidemic Final Size Datasets with Hidden Underlying Household Structure}

\author[1]{Joseph Brooks}
\author[1]{Thomas House}
\author[1]{Lorenzo Pellis}
\author[1,2]{Joe Hilton}
\affil[1]{Department of Mathematics, University of Manchester, Manchester, United Kingdom }
\affil[2]{Manchester Centre for Health Economics, University of Manchester, Manchester, United Kingdom }

\date{}

\begin{document}

\maketitle


\pagebreak

\begin{abstract}
 Households represent a key unit of interest in infectious disease epidemiology, in both empirical studies and mathematical modelling. The within-household transmission potential of an infectious disease is often summarised in terms of a secondary attack ratio, the proportion of an index case's household contacts who become infected during the index case's infectious period. Despite its widespread use, the secondary attack ratio depends on the distribution of household compositions seen during the study period, making the information it conveys difficult to generalise to new populations and different contexts. Extending estimates of transmission potential to new populations instead requires estimates of person-to-person transmission rates which can be convoluted with data on population structure to parametrise mechanistic transmission models.  

In this study we present a new Bayesian inference method which uses an MCMC algorithm to infer the transmission intensity by imputing the unreported household structure underlying the epidemic dynamics. This method can be run on household epidemiological data reported at varying levels of resolution. For synthetic data from a realistic underlying household size distribution, we were able to achieve over 90\% coverage in our estimates of transmission rate over a range of transmission intensity scenarios. The method was also able to consistently achieve over 90\% coverage for data generated with a pathological underlying household size distribution, given strong information about the household size distribution. Using an existing dataset which recorded micro-scale household epidemiological outcomes during the COVID-19 pandemic, we show that at least stratifying observed secondary attack ratios by household size substantially reduces the uncertainty in transmission parameter estimates. Our findings suggest that researchers conducting household epidemiological studies can drastically improve the utility of their results to infectious disease modellers by reporting household-stratified estimates, without needing to report full micro-level transmission outcomes. These results aim to encourage the reporting of higher resolution outputs in future epidemiological field work as, in the absence of strong priors, the transmission parameters were not easily identifiable from low resolution datasets, which are commonly reported. 
\end{abstract}

\pagebreak  
\section*{Introduction}
\input{Sections/introduction}

\section*{Methods}
\input{Sections/methods}

\section*{Results}
\input{Sections/results}

\section*{Discussion}
\input{Sections/discussion}

\section*{Conclusion}
\input{Sections/conclusion}

\section*{Acknowledgments}
\input{Sections/acknowledgments}

\section*{Code availability}
All code used to implement the methods described in this paper and to generate all figures can be found in this \href{https://github.com/joebrooks6330/hh-partition-mcmc}{GitHub repository}. 

\pagebreak
\appendix
\input{Sections/appendices}

\pagebreak
\printbibliography

\pagebreak
\beginsupplement
\section*{Supplementary plots}
\input{Sections/supplement}

\end{document}

%% file: Sections/introduction.tex
Households are important units of study in infectious disease epidemiology, both as sites of prolonged intense social contact and as convenient population units for empirical studies\cite{mossong_social_2008, madewell_household_2022}. Studies that investigate within-household transmission studies are common and provide essential data on emerging pathogens, while household structure plays an extremely important role in determining the onward spread of infection. For example, during the COVID-19 pandemic contact-tracing studies showed individuals exposed to an infected individual exposed in a household setting to be much more likely to be infected than those exposed in other settings (e.g. workplace, social)\cite{li_infection_2022}. Throughout this paper contacts refer to individuals exposed to an infected case, as opposed to exposure events of which an individual may have multiple. 
 Household structure also plays a key role in interventions against infection, with non-pharmaceutical interventions (NPIs) such as stay-at-home orders acting at the household level. Likewise, studies on determinants of vaccine uptake reveal the influence of both strictly household-level factors such as income and factors such as ethnicity which tend to correlate at the household level, as well as behavioural influences from of social contacts, of which household members form an important subgroup\cite{endrich_influenza_2009, wang_multilevel_2022, aw_covid-19_2021, schmid_barriers_2017, malik_determinants_2020, viswanath_individual_2021}. In this paper, we only consider ``final size'' datasets which report the infectious status of individuals following the conclusion of a single outbreak household. This is distinct from final size studies that may consider the infectious status based on serology after a full wave which may include infections and reinfections that result from multiple introductions. \\

In field studies of household transmission, secondary attack rate (SAR), defined as the number of secondary cases divided by the number of contacts, is ubiquitous as a measure of transmissibility of a pathogen in enclosed settings such as households. As a simple ratio, it can be calculated from relatively coarse epidemiological data, requiring field researchers only to identify cohabitants of an index case and record their infection status over a sufficiently long period. Despite its widespread use, its simplicity makes the SAR poorly suited to capture the complex dynamics of person-to-person transmission.\\

Infectious disease dynamics arise from a combination of biological and sociological factors which the SAR summarises as a single number, meaning that an estimate of SAR from an outbreak in one population provides very limited insight into the behaviour of future outbreaks in other populations with differing demographic and social contact structures. Instead, mechanistic models that are parameterised to reflect the social structure of specific populations are necessary to analyse infectious disease outbreaks and to make predictions. Mechanistic mathematical models play an important role providing quantitative evidence and models have been developed that take account the geographical, gender/sex, age, workplace and household structured complexity in social mixing patterns \cite{ball_reproduction_2016,pellis_epidemic_2011,hilton_computational_2022}
. \\

Despite the considerable literature on household-structured epidemic modelling and due to the ubiquity of the SAR, empirical household transmission studies most often report their data with SAR and logistic regression in mind. These studies often only report the number of cases and the number of contacts, giving little to no information on the household size distribution and only reporting the mean of the household final size distribution. Because each secondary infectious cases increases the infectious pressure on any remaining susceptible household members and thus increases their likelihood of being infected, the household final size distribution is often bimodal and so particularly poorly-described by its mean (see, e.g.\ \textcite{house_inferring_2022}, Figure 2). While efficient likelihood-based methods for parameterising mechanistic models exist and are relatively efficient in populations of the small scale of a household \cite{addy_final_nodate,house_inferring_2022}, they require higher resolution data than just the reported overall SAR. Therefore, in order to estimate person-to-person transmission rates from these studies we must augment the datasets, imputing the household structure that underlies them.\\

Data augmentation and data imputation have a long and successful history of use in inference for a variety of applications in infectious disease modelling \cite{oneill_bayesian_1999,jewell_bayesian_2009,worby_reconstructing_2016,touloupou_bayesian_2020}. The central problem these methods seek to address is that epidemics are rarely fully observed, meaning that epidemiological datasets are often incomplete and so we cannot directly evaluate the model likelihood based on these data alone. Data imputation approaches address this by augmenting these datasets with the missing information and treating the unobserved data as parameters. Standard parameter inference methods, often using MCMC, can then be used with additional steps which propose values for the missing data and identify scenarios which are most consistent with the outcomes we do observe.\\

In this study we present a novel mass imputation MCMC method for estimating transmission rates from household-stratified datasets at three different levels of detail: low information datasets which report the total number of contacts, the total number of secondary cases, and the number of households, with no information on how infection stratifies by household size; medium information datasets which report these same outcomes stratified by household size; and high information datasets which report the number of contacts and cases for each individual household included in the study. In the high information setting our method resembles a standard likelihood-based MCMC approach, while in the low and medium information cases we impute the number of contacts and cases in each household and evaluate the joint likelihoods of specific combinations of parameter and distribution of contacts and cases by household. Our method can incorporate existing knowledge of background demography by using the household size distribution of the ambient population as a basis for its proposals of household-specific contact and case numbers.\\

We validate our method by demonstrating that it is able to successfully recover transmission rates from synthetic data generated with an underlying household size distribution that is either taken from the UK labour force survey\cite{ons_families_2024} or a pathological “split” household so long as an accurate prior is provided for the household size distribution. When inaccurate priors on household size distribution were specified, there was a bias in the estimates of transmission rates. We demonstrate the practical applicability of our approach by estimating parameters from household-based COVID-19 studies. We use our high-information method to estimate the transmission rate and density mixing parameter from \textcite{carazo_characterization_2021}, which provides a high information set, and compare our estimates when we aggregate the data and apply our medium- and low-information methods. We find that in the low information case, the most common level of detail in household studies, there are identifiability issues. We find that contacts and cases stratified by household size are necessary to identify both parameters and that uncertainty could be reduced with datasets which report the size and number of secondary cases of each household in the study. We further estimate transmission rates from a number of low information datasets taken from a systematic review of household transmission of SARS-CoV-2 \cite{madewell_household_2022}. In this case, when only a transmission rate is being estimated, information beyond the SAR gives limited gain due to the biunivocal link between SAR and transmission rates. However, this requires strong assumptions on the relationship between household size and transmission rates as well as the distribution of infectious periods. \\

These results show that low-information datasets, which are commonly reported in household transmission studies, are insufficient to be able to parametrise the simplest of mechanistic models. Contacts and cases stratified by household size can resolve identifiability issues and datasets reporting the full outcomes of household outbreaks can further reduce uncertainty in parameter estimates.

%% file: Sections/methods.tex
\subsection*{Transmission model}
The datasets of interest in this analysis come from household transmission studies, with our transmission model simulating the processes which generated the data in a given study. We assume that each household in the study experiences an outbreak seeded by a single primary case with all subsequent infections resulting from within-household transmission, and that the maximum household size is $m+1$. The infectious period of an individual, denoted $I$, is a random variable which follows a specified distribution which is the same for all individuals. During this infectious period, individuals make infectious contact with each other individual in their household according to a Poisson process with a rate given by the household size-dependent parametric form
\begin{equation}\label{eq: transmission rate}
    \beta_n = \frac{\beta}{n^\eta} \qquad \text{for } n=1,\dots,m,
\end{equation}
where $n$ is the number of initially susceptible individuals (one less than the size of the household), hereafter referred to as \emph{contacts}, $\beta>0$ is the base transmission rate and $\eta\in[0,1]$ is a mixing parameter. This particular parametric form for a transmission rate dependent on the size of the household is often considered \cite{antonovics_generalized_1995,cauchemez_bayesian_2004,tsang_effect_2023} and allows us to move between two common mixing assumptions: density dependence ($\eta=0$) and frequency dependence ($\eta=1$) \cite{de_jong_how_1995,mccallum_how_2001,rhodes_contact_2008}. Throughout this paper we assume that
\begin{equation}\label{eq: Inf Period}
    I  \sim \text{Gamma}(a,a)
\end{equation}
so that $\mathbb{E}[I]=1$ and the expected number of secondary cases produced by a primary case in a household with $n$ initial susceptible contacts is $\beta_n$. Throughout this paper we make the choice of $a=2$. In the Supplementary material we provide results for two illustrative cases: $a =1$, which gives $I \sim \text{Exponential(1)}$ and thus Markovian infectious disease dynamics, and the limit for $a\rightarrow \infty$, which results in a constant infectious period $I  \equiv 1$.

\subsection*{Datasets}
Depending on how results from such studies are reported,  datasets can vary in the amount of information that is reported. In this paper we only consider ``final size'' data, i.e.\ we ignore information about when individuals become infected and focus on how many initially susceptible individuals, named \emph{contacts}, are recorded as having been infected by the end of the period during which the household was observed, hence having become \emph{cases}. This type of data is easier to record and analyse than longitudinal data and is commonly used to measure the transmissibility of a pathogen. We consider three categories of final size datasets distinguished by the level of information reported,  with this information level denoted by a letter $J \in \lbrace L,M,H\rbrace$. Datasets of a given information level are denoted $D_J$. Low-information datasets ($J=L$) report the total number of contacts, cases and households in a study, so that the dataset consists of just three integers. Medium-information datasets ($J=M$) report the number of contacts and cases, stratified by the size of the household, i.e.\ three integers for each household size observed in the data, including the household size itself. High information datasets ($J=H$) report the number of households with each observed combination of the number of initially susceptible contacts and the final number of cases. In this case, data may be reported in a upper (or lower) triangular table. Given the focus is on estimating within-household transmission, we do not consider fully susceptible households (generally the situation in case-ascertainment studies) or households with only one member.

\subsection*{Likelihood}
Given a dataset $D$, we aim to estimate the posterior of the transmission  parameters $\theta = (\beta,\eta)$, $\pi(\theta \: \vert \: \mathbf{D})$, using a Markov chain Monte Carlo method. Throughout the fitting process we assume a fixed value of $a$ (either $a =1,2$ or $a\rightarrow\infty$) and so all dependence on $a$ is suppressed, including in the above posterior. We first construct a likelihood that can be evaluated using a high-information dataset.

Let $Z$ denote the final size of a household outbreak. We can use results from \textcite{ball_unified_1986} to construct a set of triangular equations which can be solved to determine the exact distribution of $Z$ conditional on $n$ and $\theta$:
\begin{equation}
    \sum_{z=0}^j  {n-z\choose j-z} \frac{\mathbb{P}\left( Z=z \: \vert \: n,\: \theta \right)}{\phi(\beta (j-n))^{1+z}} = {n\choose j} \quad j=0,1,\dots,n, \label{eq: ball}
\end{equation}
where $\phi(t) = \mathbb{E} \left[ \exp \left( tI\right) \right] = \left(1-\frac{t}{a} \right)^a$ is the moment generating function of the infectious period. Alternative methods for the computation of the final size distribution could be found in \textcite{house_how_2013}. \\

In order to derive our likelihood more easily, we define an enumeration map $f$, so that we can write out a high-information dataset as a vector of counts:

\begin{equation}
    f(n,z) = \sum_{i=1}^{n-1} (i+1) + z =\frac{(n-1)(n+2)}{2} + z
\end{equation}
This enumeration orders household outcomes first in terms of the number of contacts and secondly in terms of the number of cases among these contacts. Given households of size one are not included in the dataset, the enumeration starts counting from $f(1,0) = 0, f(1,1)=1$ and $f(2,0)=2$, which are the indices corresponding to households with 1 contact that avoids infection during the study, 1 contact which becomes a non-primary case and 2 contacts which both avoid infection respectively. The counting stops at $K = f(m,m)$, which corresponds to a household of size $m+1$  (including the primary case) in which each of the $m$ contacts are infected. We also define the element wise inverses of $f$, $n(\cdot)$ and $z(\cdot )$, such that $f(n(k),z(k))=k$. We can therefore represent a high-information dataset by a flat vector, $\mathbf{C}\in \mathbb{N}^K$ where $C_k$ is the number of households with $n(k)$ initially susceptible contacts and $z(k)$ non-primary cases.\\

Our aim is now to estimate the posterior probability of parameters $\theta$ given a high-information dataset $\mathbf{C}$, which we denote by $\pi(\theta \vert \mathbf{C})$. For some choice of $\theta$ we can define a vector of final size distributions $\mathbf{P}(\theta) \in [0,1]^K$ where $P_k(\theta) = \mathbb{P}\left(Z= z(k) \: \vert \: n(k), \: \theta \right)$. For a given $\theta$ and relatively small $m$ ($\leq 20$, the scale of a household), $\mathbf{P}(\theta)$ is straightforward and efficient to calculate by solving the linear system in Equation \eqref{eq: ball}. We assume that our study population samples household sizes from an underlying household size distribution, $\mathbf{p}=(p_1,\dots,p_m)$, where $p_n$ is the probability a randomly selected household is size $n+1$. We assume a hierarchical framework for the final size outcomes $\mathbf{C}$:
\begin{align}
    \mathbf{p} \: \vert \: \boldsymbol{\alpha} &\sim \text{Dir}(\boldsymbol{\alpha}) \\
    \mathbf{C} \: \vert \: \mathbf{p}, \theta &\sim \text{Multi}(\sum C_k, \mathbf{P} \odot_{n} \mathbf{p})
\end{align}
where $\boldsymbol{\alpha} = (\alpha_1,\dots,\alpha_m)$ are chosen hyperparameters for the Dirichlet distribution and $\odot_{n}$ denotes an element wise product where the $k^{\text{th}} $ entry of the first vector is multiplied by the $n(k)^{\text{th}}$ of the second. Note that if $\mathbf{p} = (p_1, \dots , p_m) \sim \text{Dir}(\boldsymbol{\alpha})$ then $\mathbb{E}(p_i) = \frac{\alpha_i}{\alpha_0}$ where $\alpha_0 = \sum_{i=1}^m\alpha_i$. Also, larger values of $\alpha_0$ correspond to smaller variance of $\mathbf{p}$ and so that larger $\alpha_0$ can be interpreted as representing higher confidence that the household size distribution in the study is close to $\mathbb{E}(\mathbf{p})$.
Note also that, although this is similar to a Dirichlet-Multinomial distribution, it is not the same because $\mathbf{C}$ and $\mathbf{p}$ are of differing lengths $K$ and $m$ respectively. We can derive a likelihood similar to that of a Dirichlet-Multinomial as follows:
\begin{align}
    L(\theta\: \vert \: \mathbf{C}, \boldsymbol{\alpha}) &= \pi(\mathbf{C} \: \vert \: \theta,\boldsymbol{\alpha})\\ &= \int_{S_{m-1}} \pi(\mathbf{C}\: \vert \mathbf{p}, \theta)\pi(\mathbf{p}\: \vert \boldsymbol{\alpha}) \text{d}\mathbf{p} \\ 
    &= \int_{S_{m-1}} \frac{1}{\text{B}(\mathbf{C})} \prod_{k=0}^{K} \left(P_k(\theta) p_{n(k)} \right ) ^{C_k}  \frac{1}{\text{B}(\boldsymbol{\alpha})} \prod_{n=1}^m p_n^{\alpha_n-1}\text{d}\mathbf{p} \\
    &= \frac{1}{\text{B}(\mathbf{C})B(\boldsymbol{\alpha})}\prod_{k=0}^{K} P_k(\theta) ^{C_k} \int_{S_{m-1}}  \prod_{k=0}^{K}  p_{n(k)}^{C_k} \prod_{n=1}^m p_n^{\alpha_n-1}\text{d}\mathbf{p}, 
    \end{align}
    where $S_{m-1}$ is the $m-1$ simplex. If we let $\boldsymbol{\gamma} \in \mathbb{R}^m$ be defined such that $\gamma_n := \sum_{k=f(n,0)}^{f_(n,n)} C_k + \alpha_n-1$, then we can reduce the likelihood to the following:
    \begin{align}
        \pi(\mathbf{C} \: \vert \: \theta,\boldsymbol{\alpha}) &=  \frac{1}{\text{B}(\mathbf{C})\text{B}(\boldsymbol{\alpha})}\prod_{k=0}^{K} P_k(\theta) ^{C_k} \int_{S_{n-1}}    \prod_{n=1}^m p_n^{\gamma_n} \text{d}\mathbf{q}\\
        &= \frac{\text{B}(\boldsymbol{\gamma})}{\text{B}(\mathbf{C})\text{B}(\boldsymbol{\alpha})}\prod_{k=0}^{K} P_k(\theta) ^{C_k}. 
        \label{eq: likelihood}
    \end{align}
All components of this likelihood can be efficiently calculated, making it suitable for inference methods that require many evaluations of the likelihood, such as MCMC.

\subsection*{Data Augmentation and MCMC algorithm}
In the cases that we only have a medium/low-information dataset $D_J$ ($J \in \lbrace L,M\rbrace$), we would like to derive the posterior distribution of $\theta$, $\pi(\theta \: \vert D_J, \: \boldsymbol{\alpha})$. However, we cannot directly evaluate the corresponding likelihood $L(\theta \: \vert D_J, \boldsymbol{\alpha}) = \pi(D_J\:\vert \: \theta,\boldsymbol{\alpha})$. Instead, we must augment the dataset by imputing the household structure necessary to use the likelihood in Equation \eqref{eq: likelihood}. Therefore, we target $\pi(\theta, \mathbf{C} \: \vert \: D_J ,\boldsymbol{\alpha})\propto \pi(D_J,\mathbf{C} \: \vert \: \theta,\boldsymbol{\alpha})\pi(\theta) = \pi(\mathbf{C}\: \vert \: \theta,\boldsymbol{\alpha})\pi(\theta)$ where the equality comes from the fact that $D_J$ is fully determined by $\mathbf{C}$. If we can specify an appropriate proposal distribution $q_J((\theta',\mathbf{C}')\: \vert \: (\theta,\mathbf{C}))$ then we can use a Metropolis-Hastings algorithm with acceptance probability
\begin{equation}
    A((\theta',\mathbf{C}')\: \vert \: (\theta,\mathbf{C})) = \min \left( 1, \frac{\pi(\mathbf{C}'\vert \theta',\alpha)\pi(\theta')q_J((\theta,\mathbf{C}) \vert  (\theta',\mathbf{C}'))}{\pi(\mathbf{C}\vert \theta,\alpha)\pi(\theta)q_J((\theta',\mathbf{C}') \vert (\theta,\mathbf{C}))}\right)\label{eq: acceptance probability}
\end{equation}
to sample from $\pi(\theta,\: \mathbf{C} \: \vert \: D_J, \: \boldsymbol{\alpha})$, from which we can get the marginal distribution $\pi(\theta\:  \vert \: D_J, \: \boldsymbol{\alpha})$.  Note that so long as the proposal distribution results in an aperiodic and irreducible chain, the algorithm will target the required distribution.  \\

\subsection*{Proposal Distribution}
We now need to define a proposal distribution function  $q_J: (\mathbb{R}^2 \times g_J(D_J)) \times (\mathbb{R}^2 \times g_J(D_J))\rightarrow [0, \infty)$, where $g_J(D_J)$ denotes the set of compatible high-information datasets for a given medium/low-information dataset; see Appendix \ref{appendix: formal definitions of compatible sets} for a formal definition of $g_J$. Our transitions occur on $\mathbb{R}^2 \times g_J(D_J)$, the product of a discrete space $g_J(D_J)$ with specific restrictions, and a more standard space $\mathbb{R}^2$. In order to efficiently navigate this space and avoid proposing incompatible high-information datasets, we cannot use standard choices for the proposal distribution and so need to specify our own bespoke distribution. \\

Suppose we start with $(\theta_0,\mathbf{C}_0)$. Our proposal distribution allows for one of two possible moves on each iteration of the MCMC algorithm: either a new  set of transmission parameters is proposed (and $\mathbf{C}_1 = \mathbf{C}_0$) with probability $s$, or a new household structure is proposed (and $\theta_1 =\theta_0)$ with probability $1-s$, where $s$ is a chosen hyper-parameter.\\

To propose a new set of transmission parameters, we use a 2-dimensional Gaussian proposal distribution centred on $\theta$ such that $\theta_1 \sim \mathcal{N}(\theta_0,\Sigma)$. In this study, when we fit to the synthetic data we assume a known $\eta$ and we only fit $\beta$ in which case $\Sigma = \begin{psmallmatrix} \sigma & 0 \\ 0 & 0 \end{psmallmatrix}$ for some $\sigma>0$. When fitting both parameters, there is likely to be significant correlation between them which can be exploited by estimating the value of $\Sigma$ from a short initial chain in a similar way to adaptive MCMC \cite{haario_adaptive_2001}. This allows for more efficient sampling from the targeted posterior.\\

\begin{figure}[ht!]
    \centering
    \includegraphics[width=\linewidth]{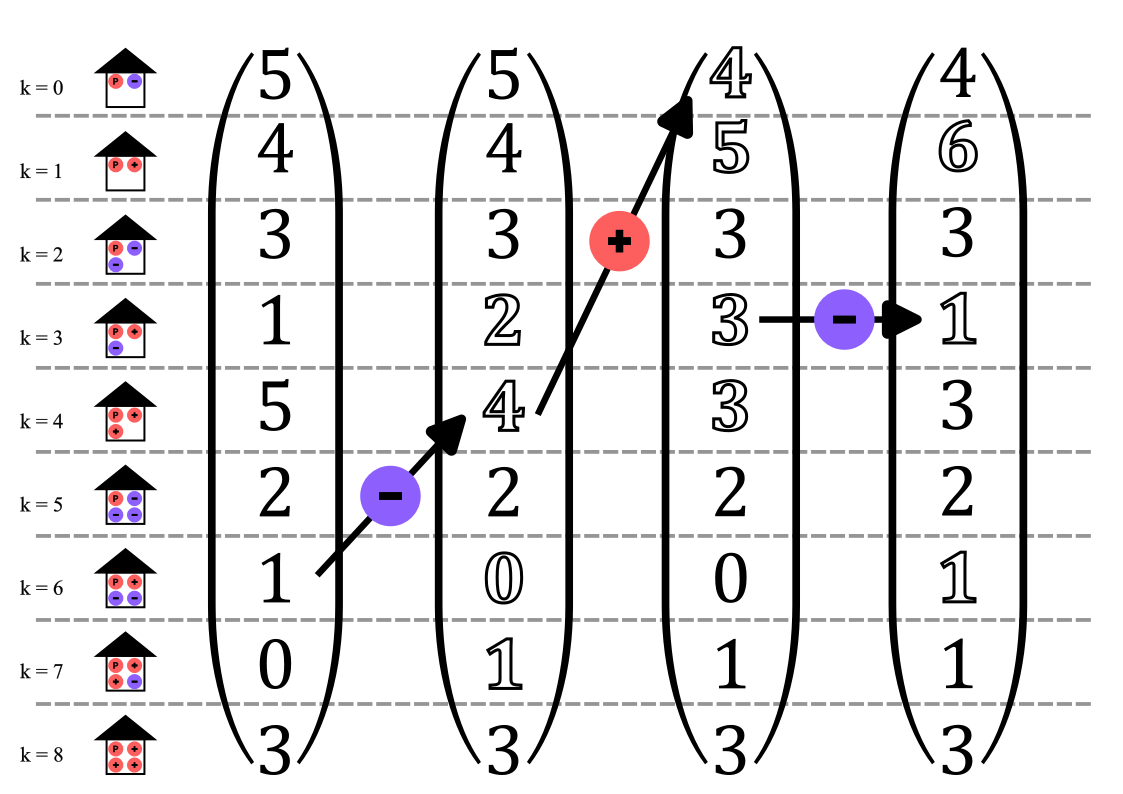}
    \caption{Visual representation of the Type 1 proposal algorithm (Algorithm \ref{alg: type 1proposal}). Three hypothetical proposal steps are shown for a low information dataset with $(N,y,n) = (24,25,45)$ and maximum number of household contacts $m=3$. On the left hand side the indices of the configurations are shown with pictorial representations of these outcomes (primary cases, secondary cases and non-cases shown by red Ps, red positive signs and blue negative signs respectively). 
    Each proposal move is represented by a black arrow starting at an entry in the previous pseudo-dataset and ending at an entry in the new pseudo-dataset with their respective indices being $k_1$ and $k_2$ selected randomly and in the notation of Algorithm \ref{alg: type 1proposal}. The infectious status of the contact being moved is shown by the symbol on the arrow. Data that have changed in a step are shown as white digits with a black outline.}
    \label{fig:algorithm_diagram}
\end{figure}

Proposing a new household structure is more complicated, requiring us to specify two distinct proposal move types for new household structures. Type 1 moves are used only in the low information case. The intuition for Type 1 moves is that individual contacts, either infected or not, are selected at random from the current household structure $\mathbf{C}_0$  and then moved from their household to another. The resulting household structure, $\mathbf{C}_1$, is still compatible with $D_L$ (i.e., $\mathbf{C}_1 \in g_L(D_L)$). This process is represented pictorially in Figure \ref{fig:algorithm_diagram} and the full algorithm is presented in Algorithm \ref{alg: type 1proposal}.\\

 Type 2 moves are used in both the low and medium information case, however there are slight differences between the two information cases. For both $J=L$ and $J=M$, Type 2 moves swap the infectious status of two individuals at random. Unlike Type 1 moves, Type 2 moves do not change the household size distribution, only the distribution of cases within households. In the case of $J=L$, households of any size can be selected for Type 2 moves. However, when $J=M$ the number of cases in households of a given size must remain fixed to ensure that $\mathbf{C}_1 \in g_M(D_M)$ and so selections are restricted to ensure households are the same size. The full algorithm for Type 2 moves is presented in Algorithm \ref{alg: type 2 proposal}.
Once a new pair $(\theta_1,\mathbf{C}_1)$ is proposed, it is accepted with the probability given in Equation \eqref{eq: acceptance probability}. \\

\subsection*{Priors}
The prior on $\theta$ in Equation \eqref{eq: acceptance probability}, $\pi(\theta)$, is the product of the two independent priors for $\beta$ and $\eta$, i.e.\ $\pi(\theta) = \pi_1(\beta)\pi_2(\eta)$. There are no restrictions on the distributions of $\pi_1$ or $\pi_2$.  Throughout this paper we do constrain $\pi_1$ and $\pi_2$ based on the meaning of the parameters: $\beta$ is a rate, so is assumed $\ge 0$, and although its range might be larger (even involving negative values), we assume $\eta\in [0,1]$. 

\subsection*{Synthetic data generation}
In order to validate our method, we generated synthetic high-information datasets with known parameters, and attempted to re-estimate $\beta$ (with $\eta$ known and fixed) from the corresponding low-information dataset. To assess the impact of sample size on fitting behaviour we generated datasets with $N=100$ and $N=1000$ households, with sizes sampled from a given household size distribution. We carried out this process with two household distributions, one informed by the 2023 UK Labour Force Survey (LFS) \cite{ons_families_2024} and another one constructed to be an extreme example in which most of the households are either of the smallest or largest possible size, i.e.\ 2 or 6 (``split'' distribution) (Figure \ref{fig: hh size distributions}). In both cases, the maximum number of contacts is $m=5$. For details of these distributions see Appendix \ref{appendix: hh size dist}. The final sizes of each outbreak are then sampled from the probabilities in Equation \eqref{eq: ball}.

\subsection*{Choice of household size distribution prior $\boldsymbol{\alpha}$}
The household size distribution prior $\boldsymbol{\alpha}$ should be chosen so that the expectation of $\text{Dir}(\boldsymbol{\alpha})$ is proportional to the household size distribution that the data is assumed to have come from. The choice of the scaling factor $\alpha_0$ controls both the speed of convergence to the specified distribution and the size of the variance (higher values of $\alpha_0$ leads to quicker convergence and smaller variance). \\

When validating the method on synthetic datasets we used choices of $\boldsymbol{\alpha}$ proportional to the household size distributions used to generate the datasets. For both the 2023 UK Labour Force Survey (LFS) and our split distribution we chose $\alpha_0 =100$. To assess the impact of misspecified distributions, we fitted each dataset using both choices of $\boldsymbol{\alpha}$, meaning each dataset was fit with an $\boldsymbol{\alpha}$ proportional to the household size distribution used to generate the data and another $\boldsymbol{\alpha}$ proportional to the other, very different, distribution. \\

When the number of households is large and the value of $\alpha_0$ small, this algorithm struggles to explore the posterior efficiently. This is because the small variance of the Dirichlet distribution causes the household structure to change slowly and so the posterior is not explored efficiently, analogously to a narrow Gaussian proposal distribution with small variance. For this reason, it is often necessary to make a strong choice about $\boldsymbol{\alpha}$, keeping the high dimensional household structure relatively similar throughout, in order to infer the transmission rate. Uninformative choices, such as a Jeffrey's prior (where $\alpha_i = 0.5$ for $i = 1,\dots m$), lead to poor mixing in both the household structure and the transmission parameter $\beta$.


%% file: Sections/results.tex


\subsection*{Validation on synthetic data}
We first validated our method, estimating the transmission parameter $\beta$ from synthetic datasets generated from known parameters using Equation \eqref{eq: ball}. To assess robustness to different transmission scenarios, we performed validation for three different values each of $\beta$ and $\eta$ and using both the UK and split household distributions, resulting in a total of 18 combinations for which we generated 100 datasets of 1000 households each. When estimating $\beta$ from each synthetic dataset we fixed $\eta$ to the value used to generate the dataset. In each case, we assessed the impact of misspecified household size distributions by repeating the inference algorithm with the parameters for the Dirichlet distribution ($\boldsymbol{\alpha}$) proportional to (i.e.\ the expectation of the Dirichlet was equal to) each of the UK household size distribution and the extreme ``split'' distribution (both with $\alpha_0 = 100$) \\

\begin{figure}
    \centering
    \includegraphics[width=\linewidth]{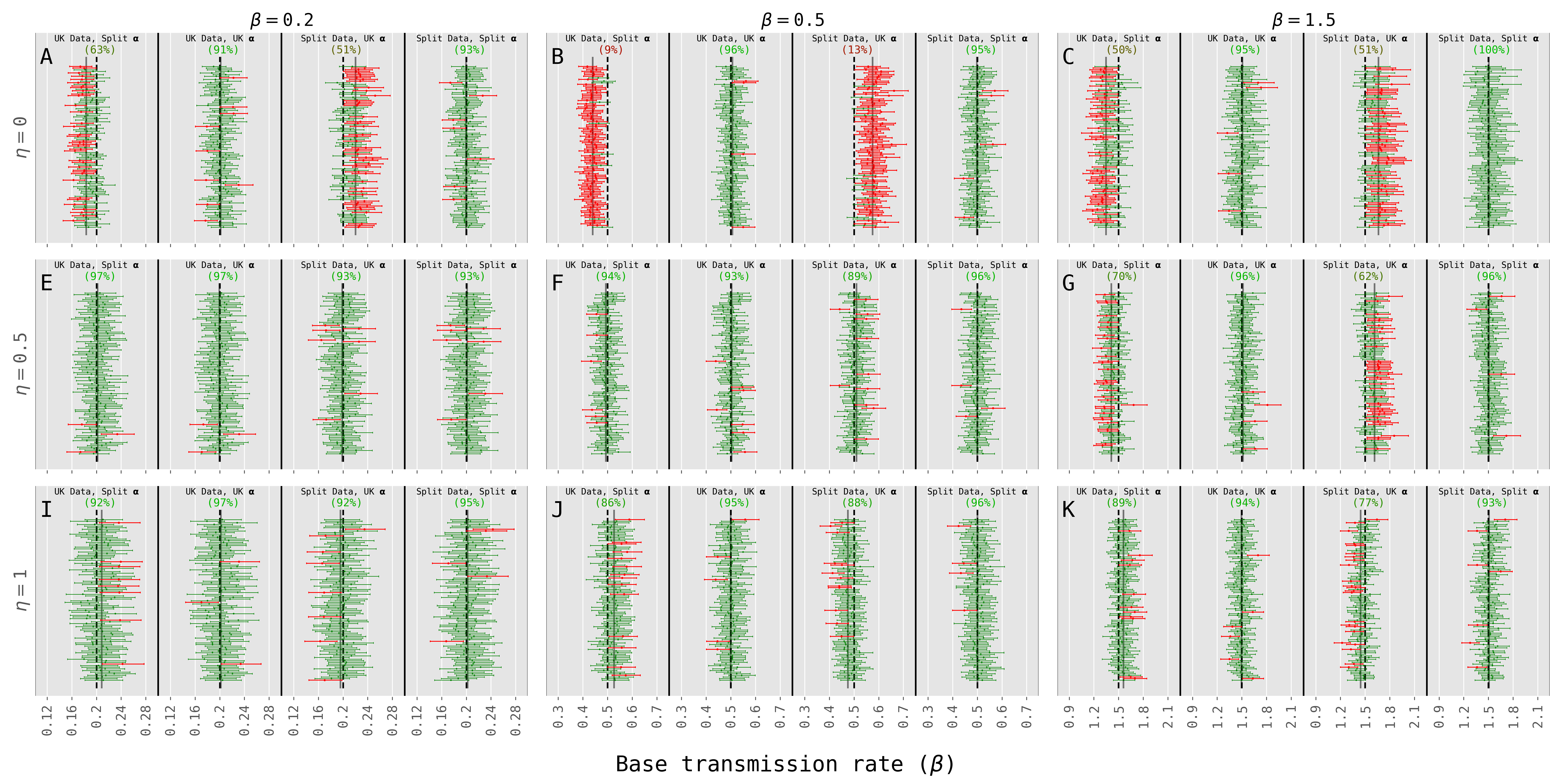}
    \caption{In each subplot, 95\% credible intervals for the posteriors of $\beta$ are shown for 100 low information synthetic datasets per household size distribution. Each dataset is generated using the Ball model ($I\sim \text{Gamma}(2,2)$) and $N=1000$ household sizes sampled either from the UK LFS (2023) or the ``split'' household size distribution (see Appendix \ref{appendix: hh size dist} for details), and $\beta$ is re-estimated separately using $\boldsymbol{\alpha}$ proportional to each of the UK LFS and split household size distributions ($\alpha_0 = 100$). Each subplot row shows results for synthetic data generated with different base transmission rate ($\beta$) and each column shows results for different density mixing parameter ($\eta$). The value of $\beta$ used to generate the data is shown by the vertical dotted line in each plot and credible intervals are plotted in green if this value is contained in them and red otherwise. The mean estimate of $\beta$ across all 100 datasets is shown by the grey vertical line. The percentage of synthetic datasets for which the real value is contained within the 95\% credible interval is shown above each set of credible intervals. All fits were done with $I\sim \text{Gamma}(2,2)$ for a known $\eta$ and so only $\beta$ was inferred.}
    \label{fig: synthetic data validation 1000}
\end{figure}

The 95\% credible intervals for $\beta$ obtained from this validation process are plotted in \ref{fig: synthetic data validation 1000}. Each panel shows the 95\% credible intervals for $\beta$ estimated from 100 synthetic datasets arranged according to the values of $\beta$ (columns) and $\eta$ (rows) used to generate the data, and, within each panel, the combination of true household size distribution used to generate the data and assumed household size distribution used to perform the inference. \\
 
When the choice of $\boldsymbol{\alpha}$ matches the household size distribution used to generate the synthetic data the correct parameters are recovered by the MCMC algorithm at least 90\% of the time for all 9 choices of parameter vector $\theta$. However, when data was generated from the UK distribution and an $\boldsymbol{\alpha}$ proportional to the ``split'' distribution was used the MCMC failed to recover the parameters for a number of values of $\theta$. When $\eta=0$ (density-dependent mixing), there was systematic underestimation; for $\theta =(0.2,0),(0.5,0)\text{ and } (1.5,0)$ only $63\%, 9\%$ and $50\%$ of the datasets result in posteriors with $95\%$ intervals containing the real value. For $\eta=0.5$ (intermediate density-dependence), estimates were less dependent on $\boldsymbol{\alpha}$ but for $\theta=(1.5,0.5)$ there was only $70\%$ due to underestimation again. However, for $\eta=1$ (frequency-dependent mixing), there was some overestimation; for $\theta=(0.2,1),(0.5,1) \text{ and } (1.5,1)$ $92\%,86\%$ and $89\%$ of datasets results in credible intervals containing the real value of $\beta$. In the converse situation, when data was generated from the ``split'' distribution and an $\boldsymbol{\alpha}$ proportional to the UK distribution was used, we saw comparably poor performance, but typically in the opposite direction. Thus, where there was overestimation in the UK data and split $\boldsymbol{\alpha}$ case there was underestimation of $\beta$ and vice-versa. \\

The same figures for datasets with $N=100$ households are shown in Figure \ref{fig:synthetic data validation 100}. Similar biases were observed for these smaller datasets but the coverage was better due to the wider credible intervals. Box-and-whisker plots showing the distribution of ESS values for datasets with $N=100$ and $1000$ households are shown in Figures \ref{fig:ESS box and whisker 100} and \ref{fig:ESS box and whisker 1000} respectively. For the medium and high-information cases we performed a more expansive analysis in which we fit both $\eta$ and $\beta$ simultaneously. We found that the MCMC method was able to reliably recover both parameters across all scenarios, as presented in Figure \ref{fig: synthetic data validation 1000} (not shown).

\subsection*{Fitting to empirical data}
\textcite{carazo_characterization_2021} reports final size data from outbreaks in households of health care workers in Qu\'ebec, Canada during the COVID-19 pandemic. In that study, final size data is reported in sufficient detail for a high information dataset. This allowed us to compare the performance of the MCMC method on a real dataset at each of the three information levels. For the choice of $\boldsymbol{\alpha}$ we used data from the 2021 Census in Canada \cite{government_of_canada_profile_2022} which reported the number of households of each size in Qu\'ebec. We use this distribution in the same way as the UK LFS (see Appendix \ref{appendix: hh size dist}); however, given the census aggregated all households with 5 or more individuals, we arbitrarily distributed these households between sizes 5 and 6 at a 2:1 ratio. For our parameter estimation we used an $\boldsymbol{\alpha}$ with expectation proportional to this distribution with $\alpha_0 =100$.\\

\begin{figure}
    \centering
    \includegraphics[width=\linewidth]{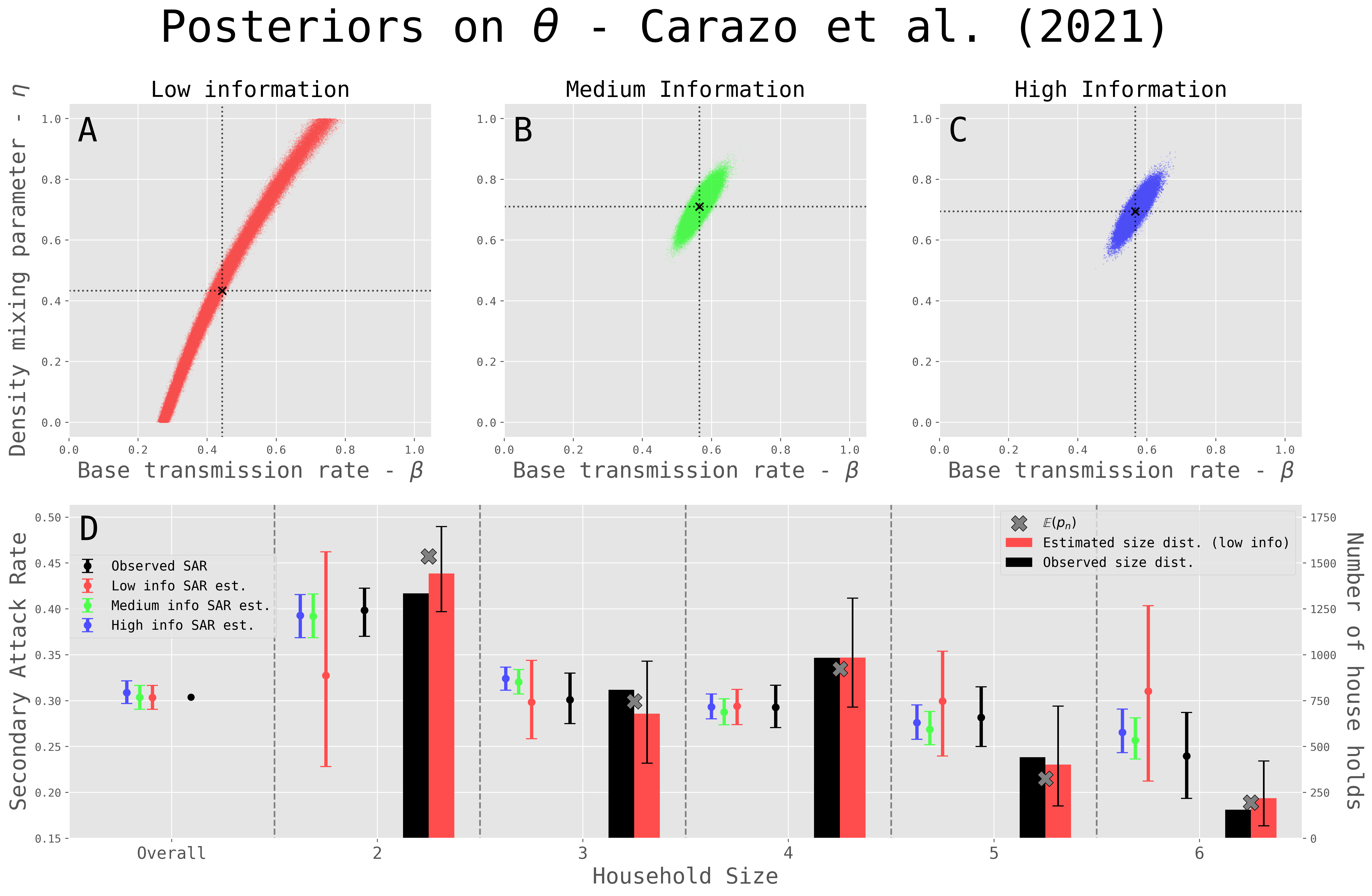}
    \caption{Panels A, B, and C show the posterior distributions of $\theta$ obtained by fitting the low- (red), medium- (green), and high-information (blue) versions of the data from \textcite{carazo_characterization_2021}, respectively. Panel D displays the secondary attack rate (SAR) implied by each posterior alongside the observed SAR (black) for each household size and overall; error bars represent 95\% credible intervals, with those for the observed SAR estimated by bootstrapping. Bars indicate the mean number of households of each size in the low-information fit (red, 95\% CI shown) compared with the observed distribution (black). Grey Xs indicate the expected number of households of each size implied by the choice of $\boldsymbol{\alpha}$. The infectious period is assumed to be $I \sim \text{Gamma}(2,2)$.}
    \label{fig: carazo Scatter Gamma2}
\end{figure}

Panels A, B and C in Figure \ref{fig: carazo Scatter Gamma2} show the posteriors obtained by simultaneously fitting both $\beta$ and $\eta$ to the low- (red), medium- (green) and high-information datasets (blue). For each of these we used a $\text{Uniform}(0,1)$ prior on $\eta$ and an improper (positive reals) prior on $\beta$. These results were obtained with $I\sim \text{Gamma}(2,2)$, and the results for $I\sim \text{Exponential}(1)$ and $I=1$ are shown in Supplementary Figures \ref{fig: carazo Scatter Markov} and \ref{fig: carazo Scatter Fixed} respectively. Panel A of Figure \ref{fig: carazo Scatter Gamma2} indicates that with low-information data there was poor identifiability between $\beta$ and $\eta$, with the MCMC exploring all values of $\eta$ (limited by assumption to [0,1]) and producing a strip of values compatible with the observed SAR. With the medium-information dataset, instead, the parameters were clearly identifiable and the posterior was only slightly wider than that obtained based on the high-information dataset. The means in these two cases -- indicated by the black crosses -- are 0.57 (0.52, 0.63) for $\beta$ and  0.71 (0.62, 0.80) for $\eta$ given the medium-information dataset, and 0.57 (0.52, 0.62) for $\beta$ and 0.69 (0.61, 0.78) for $\eta$ given the high information one.\\

The error bars in Panel D of Figure \ref{fig: carazo Scatter Gamma2} indicate the estimated overall and size-stratified SAR for each of the information levels and in comparison with the observed values from \textcite{carazo_characterization_2021} (black dots). The low-information credible intervals appear to cover the observed value for all sizes but are very wide, particularly for households of size 2 or 6. Adding higher levels of information incrementally narrows the credible intervals and the medium- and high-information SAR estimates are close to those observed in the data. The bar chart in Panel D of Figure \ref{fig: carazo Scatter Gamma2} shows the number of households of each size from the low information fit (red) and the observed data (black). The error bars cover the observed household size distribution, showing that the assumption that households in this survey were sampled from the size biased distribution of household sizes in Qu\'ebec was well founded.\\

To assess the behaviour of our method for data from a wider range of settings, we estimated the transmission rate for SARS-CoV-2 based on data from 23 papers \cite{gomaa_incidence_2021,ng_risk_2022,watanapokasin_transmissibility_2021,hsu_household_2021,lewis_household_2020,verberk_transmission_2022,baker_sars-cov-2_2022,jalali_increased_2022,ng_impact_2022,gorgels_increased_2022,tanaka_increased_2021,teherani_burden_2020,sundar_low_2021,wang_basic_2020,dub_high_2022,jashaninejad_transmission_2021,carazo_characterization_2021,ogata_increased_2022,remon-berrade_risk_2021} (25 datasets) that reported a low-information dataset, taken from a list of papers identified in a series of literature reviews by Madewell \emph{et al.} \cite{madewell_household_2020,madewell_factors_2021,madewell_household_2022}. For each of these low information datasets we fit the base transmission rate ($\beta$) and mixing parameter ($\eta$). We used the posterior obtained from \textcite{carazo_characterization_2021} (Panel C, Figure \ref{fig: carazo Scatter Gamma2}), combined with an improper prior on $\beta$, as the prior, i.e.\ $\pi(\theta) = \pi_2(\eta)$. Using this prior was necessary given the issues with identifiability found for the low information case in Figure \ref{fig: carazo Scatter Gamma2}, but the resulting MCMC chains mixed well, as shown in the trace plots in Figures \ref{fig:mw traceplot1}, \ref{fig:mw traceplot2} and \ref{fig:mw traceplot3}. We again assuming $I\sim \text{Gamma}(2,2)$. For each study we found census or similar data to inform $\boldsymbol{\alpha}$ (in all cases, the data was size-weighted as discussed in Appendix \ref{appendix: hh size dist} and we used $\alpha_0 = 200$). Each low information dataset and source for $\boldsymbol{\alpha}$ can be found in the supplementary CSV file.  Figure \ref{fig:MW_beta_fits} shows the estimates and 95\% credible intervals for $\beta$ based on each of these datasets, split by variant (where specified) and colour-coded accordingly. The table on the right hand size shows the reported SAR, mean household size (including index cases) and number of households ($N$) for each study.\\

\begin{figure}
    \centering
    \includegraphics[width=\linewidth]{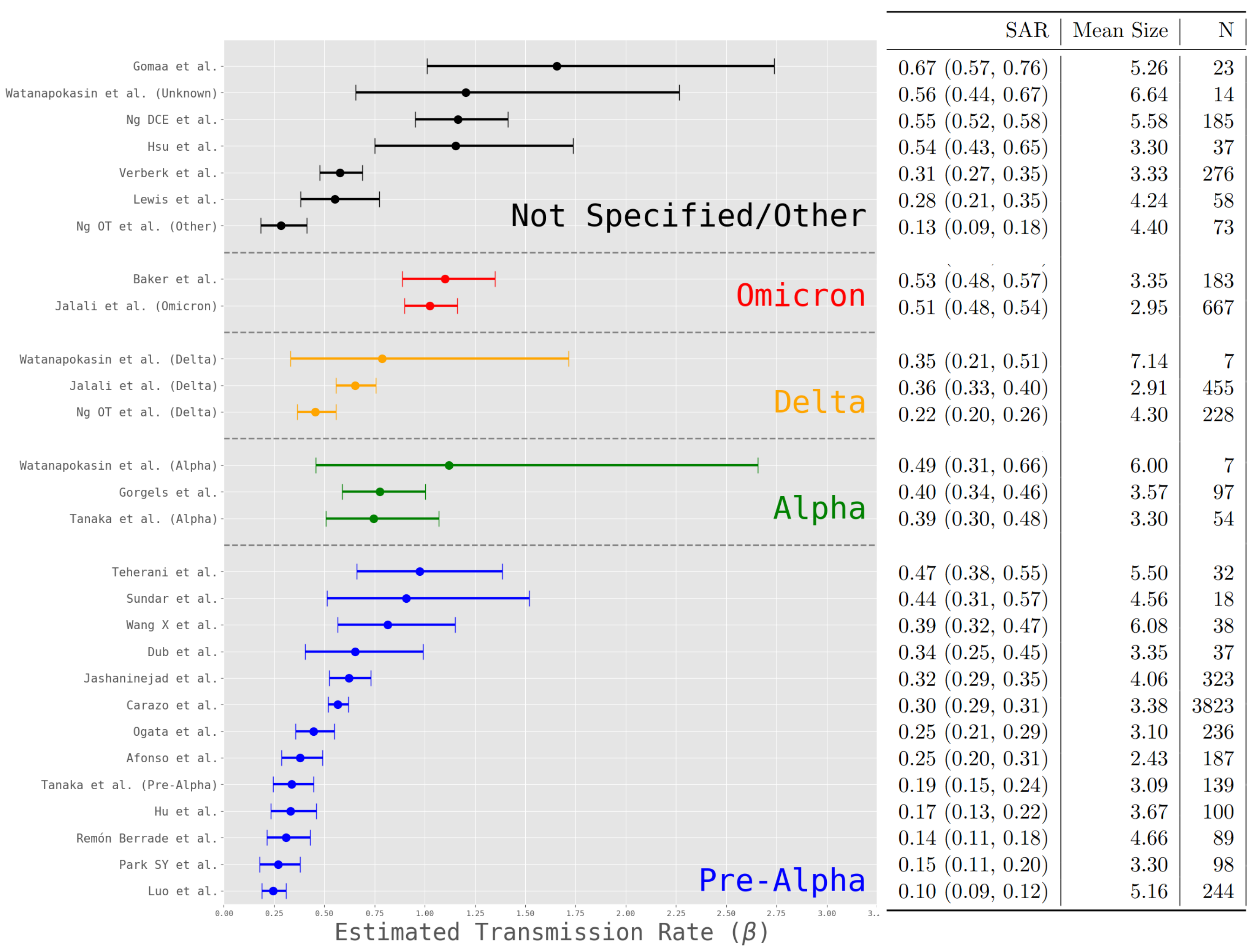}
    \caption{
    Base transmission rates ($\beta$) for SARS-CoV-2 estimated from low-information datasets from studies included in Madewell \emph{et al.} \cite{madewell_household_2020,madewell_factors_2021,madewell_household_2022}, colour coded and grouped by strain where it was specified. A table on the right hand side lists the SAR (binomial 95\% CIs), average household size and number of households in each of these studies.}
    \label{fig:MW_beta_fits}
\end{figure}

We successfully inferred transmission parameters from low-information datasets, combining prior information on $\eta$ (in this case obtained from the high-information dataset in \textcite{carazo_characterization_2021}) and information on the household size distribution of the relevant populations in $\boldsymbol{\alpha}$. We were able to properly account for differences in household size and better quantify the uncertainty compared to binomial confidence intervals often used for SARs.

%% file: Sections/discussion.tex
In this study we have introduced a novel parameter estimation algorithm designed to address the  ubiquity of coarse ``low information'' datasets reported in household final size studies. Our approach augments likelihood-based MCMC by jointly imputing the distribution of household sizes in the study population and the distribution of secondary cases across these households. In developing this approach our aim is to contribute a new tool for researchers parameterising  mechanistic models based on household final size data, while our exploration of the algorithm's behaviour given synthetic and empirical data demonstrates both the capabilities of our method and the relative values of low-, medium-, and high-information datasets as outputs from household final size studies. Further to this, enhances the understanding of the dependence of SAR on household size distribution, transmission rates and mixing assumptions like frequency dependence. \\

When an accurate prior on the household size distribution was specified, the algorithm reliably recovered the transmission rate $\beta$ for datasets generated from both a realistic household size distribution -- in this paper, from the UK Labour Force Survey (UK LFS) -- or an unrealistic ``split'' household size distribution, for a range of parameter values of both $\beta$ and $\eta$. For all 9 parameter combinations we considered, the true value of $\beta$ was covered by the 95\% credible interval. However, when the prior didn't correspond to the household size data used to generate the data, there was often systematic bias in the estimates of $\beta$, particularly under an assumption of density dependent mixing ($\eta=0$).\\

Given the strong non-linear relationship between the transmission rate and the SAR for different household sizes (see Figure \ref{fig:FS/SAR supplement}), the same value of $\beta$ leads to a different SAR from the ``split'' distribution compared to the UK LFS distribution. In particular, for $\eta=0$ (bottom left panel of Figure \ref{fig:FS/SAR supplement}), the former is larger than the latter. Therefore, if data is generated from the ``split'' generation but the prior is chosen to be proportional to the UK LFS distribution, a significantly larger $\beta$ is required to fit the SAR, causing the observed overestimation. This discrepancy is strongest for values of $\beta$ around 0.5, in line with what can be seen in Figure \ref{fig: synthetic data validation 1000}. In the reverse scenario, where data is generated from the UK LFS distribution and the ``split'' distribution is used as a prior, the same bias is observed but reversed.\\
The same qualitative discrepancy is seen for $\eta = 0.5$, though to a much less extent and reaching its maximum around 1.5, resulting in consistent overestimation for $\beta=1.5$ and $\beta=2.0$. The discrepancy is minimal for $\eta=1$, and interestingly is slightly reversed, which is compatible with the slight underestimation observed in this case in Figure \ref{fig: synthetic data validation 1000}.\\

We then investigated the difference in performance when estimating both $\beta$ and $\eta$ simultaneously from low-, medium- and high-information datasets from \textcite{carazo_characterization_2021}. The resulting posteriors on ($\beta,\eta$) demonstrate that at least medium-information datasets are necessary to confidently identify both simultaneously and that simply reporting low-information datasets is insufficient for transmission parameters to be estimated with any confidence in the absence of informative priors. However, if detailed data on the final size of outbreaks in each household cannot be reported in full, then total contacts and cases stratified by household size could still be used to identify both the transmission rate and the density mixing parameter almost to the same level of accuracy. \\

The posteriors resulting from medium and high information datasets appeared very similar with little to no uncertainty reduction from using the high information dataset. This is likely due to our model choice, in particular assuming the parametric form of $\beta_n$ (Equation \ref{eq: transmission rate}) and assuming the distribution of the infectious period (Equation \ref{eq: Inf Period}). If we had instead assumed a non-parametric form of $\beta_n$, with a separate transmission rate for each household size \cite{cauchemez_simon_household_2009}, both the medium and high information datasets would likely be able to identify these parameters. If we had further allowed the parameter $a$ used to define the infectious period distribution in Equation \ref{eq: Inf Period} then the medium information would likely experience identifiability issues which could be remedied by the high information dataset. \\ 

The medium and high information estimates of SAR by household size in Figure \ref{fig: carazo Scatter Gamma2} Panel D decline gradually with household size, in line with he power law assumption on $\beta_n$. However, this does not necessarily match the values observed in the data. This may point to a misspecification of the model and motivate increasing the complexity of the model by fitting independent transmission rates for each household size.

Finally, we estimated $\beta$ for a number of household studies which provided at least low-information datasets taken from the Madewell \emph{et al.} systematic reviews \cite{madewell_household_2020,madewell_factors_2021,madewell_household_2022}. Due to differences in mean household sizes, the rank of a studies SAR does not always match the rank of the estimated transmission rate, see \textcite{park_coronavirus_2020} and \textcite{remon-berrade_risk_2021} but in general the rank is the same suggesting that in this case SAR is a reasonable single number summary of transmission potential in this case. However, it should be noted that the prior used from \textcite{carazo_characterization_2021} with $\eta \approx 0.7$ is a value for which the dependence on household is smaller (see Figure \ref{fig:FS/SAR supplement}). \\

 Only 25 of the 144 studies included in these reviews a) tried to identify co-primary cases, b) dealt with suspected co-primaries in a way that preserved the household sizes  and c) reported sufficient detail so that a low-information dataset could be extracted or derived. The inclusion criteria for \textcite{madewell_household_2022} included the ability to calculate a SAR from the reported data. This demonstrates that the vast majority of studies report their data with SAR in mind and do not report any more detail than would be necessary for this calculation.\\

Using existing inference methods, we would not be able to estimate transmission rates from these datasets. However, the framework presented in this paper allows us to consolidate these low-information final size datasets with household size data and prior information on the density mixing parameter to estimate transmission rates. Compared to the Binomial confidence intervals often used for estimates of SAR, the error bars in Figure \ref{fig:MW_beta_fits} more accurately show the uncertainty in our estimates. \\

In order to fit a transmission model to low-information data we were required to keep make simplifying assumptions such as the parametric form of $\beta_n$ and the specified infectious period distribution. Another limitation of the model comes from our assumption of a single primary case. Many household studies limit the period of follow up in order to limit the risk of multiple independent introductions. However,  multiple primary cases are not uncommon, e.g.\ when multiple residents attend social events together and so would be at risk of contracting a disease from a shared source.\\

One of the common uses of household-stratified transmission surveys is to estimate the relative susceptibility and infectivity associated with characteristics such as age, sex and ethnicity. In this work, however, we limited ourselves to identical individuals, all mixing homogeneously within each household, both to simplify the model for testing and to ensure that $g_J(D_J)$ was not too large. However, future work could relax this assumption and perform a similar investigation to estimate the effects of risk factors on transmission.\\

The choice of household structure proposal distributions was made to keep (reverse) proposal probabilities easily calculable. However, Type 1 and Type 2 moves only make small changes to the household structure and so poor mixing can occur, particularly when datasets are large (1000s of households) and strong priors on the household size distribution are not specified. Further consideration of alternative proposal distributions, for example uniform sampling from $g_J(D_J)$, could be considered in further work, although efficient sampling algorithms from discrete and high dimensional spaces are difficult to construct. 

These results demonstrate the importance of the size distribution of the household study, which are often not reported in low-information datasets. In the absence of that information, assumptions need to be made about such a distribution and the confidence we have in it. However, such assumptions interact in non-trivial ways with the amount of person-to-person transmission and the assumed way in which transmission scales with the household size. However, although discrepancies emerge particularly strongly when mixing is close to being density-dependent ($\eta$ close to 0), for respiratory diseases such as SARS-CoV-2, which is the example used in this paper, it is common to assume frequency dependent mixing ($\eta=1$). In this case, estimates appear to be much less affected by deviations from the true distribution of the household size in the studies the low-information dataset comes from. \\

%% file: Sections/conclusion.tex
In this paper we presented an MCMC algorithm that can estimate transmission rates from very coarse datasets of the type commonly reported in household transmission studies. We successfully recovered parameters from synthetic datasets and estimated transmission rates for COVID-19 under an assumption of frequency-dependent mixing from a number of household studies. \\

However, we also demonstrated that, in the absence of prior information on the parameter controlling how person-to-person transmission varies with household size, low information datasets are generally insufficient to resolve both this mixing parameter and the transmission rate. This result suggests researchers conducting household transmission studies can greatly improve the usefulness of their results for parametrising mechanistic models by reporting contacts and cases stratified by household size and, if possible, full information on the frequency of each outcome. 

%% file: Sections/acknowledgments.tex
TH, LP, and JH are supported by the Wellcome Trust (Grant Number 227438/Z/23/Z). TH and LP are also supported by the Medical Research Council (Grant Number UKRI483). JB is an affiliate, JH is a fellow and TH and LP are members of the JUNIPER partnership , which is supported by the Medical Research Council (grant number MR/X018598/1).

%% file: Sections/appendices.tex
\section{Formal Definitions of $g_J(D_J)$}\label{appendix: formal definitions of compatible sets}

\subsubsection*{Medium-Information Datasets}
Let a medium information dataset be encoded in a tuple of two vectors $D_M = (\mathbf{n},\mathbf{z}) \in \mathbb{N}^m \times\mathbb{N}^m$ where $\mathbf{n}\geq\mathbf{z} $ element-wise. We can then define a function $g_M: \mathbb{N}^m \times  \mathbb{N}^m \rightarrow \mathcal{P}(\mathbb{N}^K)$ (where $\mathcal{P}(S)$ denotes the power set of of a set $S$) such that $g_M(D_M)$ is the set of high information datasets compatible with the medium information dataset $D_M$. For the purpose of a formal definition of $g_M$ we define two matrices $\mathbf{A},\mathbf{B} \in \mathbb{N}^{m\times K}$ where
\begin{align}
    A_{i,k} &=  
\begin{cases} i  \quad \quad \quad \quad &\text{if } f(i,0)\leq k \leq f(i,i)\\
			  0 &  \text{otherwise}
\end{cases} \\
B_{i,k} &=  
\begin{cases} k - f(i,0) & \: \text{if } f(i,0)\leq k \leq f(i,i)\\
			  0 & \: \text{otherwise}
\end{cases}
\end{align}
Then 
\begin{equation}
    g_M((\mathbf{n},\mathbf{z})) = \lbrace \mathbf{C} \in \mathbb{N}^K: (\mathbf{AC} = \mathbf{n})\:  \land \: (\mathbf{BC} = \mathbf{z} )\rbrace
\end{equation}
Similar to the high-information case, since $\sum_{k=f(n,0)}^{f(n,n)}C_k$ is the same for all $n=1,\dots,m$ and $\mathbf{C} \in g_M((\mathbf{n},\mathbf{z}))$ the choice of $\boldsymbol{\alpha}$ does not effect the likelihood.

\subsubsection*{Low-information Datasets}
Let a low-information dataset be encoded by three integers in a 3-tuple, denoted  $D_L = (N,n,z)\in \mathbb{N}^3$ where $z \leq n$ and $N \leq n \leq mN $. $N$ denotes the number of households, $n$ the total number of contacts and $z$ the total number of cases. We can then again define a function $g_L:\mathbb{N}^3 \rightarrow \mathcal{P}(\mathbb{N}^K)$ where $g_L(D_L)$ is the set of high information datasets compatible with the low information datasets.  For low information datasets we will define $g_L(D_L)$ similarly to how we did for $g_M(D_M)$. We define two vectors, $\mathbf{w}$ and $\mathbf{v} \in \mathbb{N}^{K}$, such that $w_k = n(k)$ and $v_k = z(k)$ and so

\begin{equation}
    g_L((N,n,z)) = \lbrace \mathbf{C} \in \mathbb{N}^K: (\langle \mathbf{C}, \mathbf{w} \rangle = n) \: \land  \: (\langle \mathbf{C}, \mathbf{v} \rangle = z) \: \land  \: (\langle \mathbf{C}, \mathbf{1} \rangle = N)  \rbrace 
\end{equation}
\pagebreak

\section{Household structure proposal algorithms}
\begin{algorithm}[p!]
\caption{Algorithm for Type 1 proposals of new household structures. Only used for the low information case.}\label{alg: type 1proposal}
    \begin{algorithmic}[1]
        \Require $\mathbf{C} = (c_0,\dots, c_{K})$
        \State Calculate the total number of households with at least 2 contacts,
        \[S_1 = \sum_{\{k : n(k) \geq 2\}} C_k.\]
        \State Calculate a probability distribution proportional to the number of households with each outcome and conditioned on there being at least 2 contacts $P_1 = (0,0,\frac{C_2}{S_1},\dots,\frac{C_{K}}{S_1})$
        \State Sample an index $k_1$ from $P_1$ and denote $n(k_1)$ and $z(k_1)$ by $n_1$ and $z_1$ respectively. 
        \State Uniformly draw a random number $u_{\text{inf}}\in[0,1]$
        \State \textbf{if} {$u_{\text{inf}} \leq \frac{z_1}{n_1}$} \textbf{then} an infected contact is chosen and we set $s= 1$ \textbf{else} set $s=0$ 
        \State Calculate the total number of households with fewer than $m$ contacts, 
        \[
        S_2 = \sum_{\{k : n(k) \leq (m-1)\}} c_k
        \]
        \State Calculate another probability vector $P_2$ conditioned on there being fewer than the maximum number of contacts such that for $k=0,\dots,K$:
        \begin{equation}
        P_{2k} = 
            \begin{cases}
                \frac{c_k}{S_2-1} &\text{ if } n(k)<m \text{ and } k\neq k_1 \\
                \frac{c_k-1}{S_2-1}&\text{ if } k=k_1 \\
                0 & \text{ otherwise}
            \end{cases}
        \end{equation}
        \State Draw a second index, $k_2$, from $P_2$ and denote $n(k_2)$ and $z(k_2)$ by $n_2$ and $z_2$ respectively.
        \State Define a new frequency count partition $C^* = (c_0^*,\dots,c_{K}^*)$ by starting with a copy of the original frequency count partition, $\mathcal{C}$. 
        \State Update the entries $c_{k_1}^*$ and $c_{k_2}^*$ in turn by subtracting 1. Note it could be that $k_1=k_2$ in which case we subtract 2 from $c_{k_1}^*$ in this step.
        \State Define 
        \[
        k_3 = f(n_1-1,z_1-s) \text{ and } k_4 = f(n_2+1,z_2+s).
        \]
        \State Update the entries $c_{k_3}^*$ and $c_{k_4}^*$ in turn by adding 1. Note that it is possible that $k_3=k_2$ and/or $k_4=k_1$ and the entries at those indices are unchanged.
    \end{algorithmic}
\end{algorithm} 

\begin{algorithm}[h!]
\caption{Algorithm for Type 2 proposals of new household structures. Used in both the low and medium information cases.}\label{alg: type 2 proposal}
    \begin{algorithmic}[1]
        \Require $C = (c_0,\dots, c_{K})$, $J\in \lbrace L,M\rbrace$ 
        \If{$J=L$}
            \State $m_1 =1$, since there are no infected individuals if $k=0$
        \ElsIf{$J=M$}
            \State $m_1=2$, since moving between households of 1 contact doesn't change the structure
        \EndIf
        \State $S_1 = \sum_{k=m_{1}}^{K} c_kz(k)$
        \State Calculate a probability distribution, $\mathbf{P}_1$, proportional to total number of cases in households of each outcome. For $k=0,\dots,K$:
        \[
        P_{1k} =  \begin{cases}
                \frac{c_kz(k)}{S_1} &\text{ if } m_1\leq k \leq K \\
                0 & \text{ otherwise}
            \end{cases}
        \]
        \State Sample an index $k_1$ from $P_1$. Denote $n(k_1)$ and $z(k_1)$ by $n_1$ and $z_1$ respectively.
        \If{$J=L$}
            \State $m_2 =0$, $M_2=K-1$
        \ElsIf{$J=M$}
            \State $m_2=f(n_1,0)$, $M_2 = f(n_1,n_1)$
        \EndIf
        \State $S_2 = \sum_{k=m_2}^{M_2}(n(k)-z(k))c_k$
        \State Calculate another probability vector $P_2$. For $k=0,\dots,K$:
        \[
        P_{2k} = 
            \begin{cases}
                \frac{c_k(n(k)-z(k))}{S_2-1} &\text{ if } m_2\leq k \leq M_2 \text{ and } k\neq k_1\\
                \frac{(c_k-1)(n(k)-z(k))}{S_2-1}&\text{ if } k=k_1 \\
                0 & \text{ otherwise}
            \end{cases}
        \]
        \State Draw a second index, $k_2$, from $P_2$
        \State Define a new frequency count partition $C^* = (c_0^*,\dots,c_{K}^*)$ by starting with a copy of the original frequency count partition, $\mathcal{C}$. 
        \State Update the entries $c_{k_1}^*$ and $c_{k_2}^*$ in turn by subtracting 1. Note it could be that $k_1=k_2$ in which case we subtract 2 from $c_{k_1}^*$ in this step.
        \State Update the entries $c_{k_1-1}^*$ and $c_{k_2+1}^*$ in turn by adding 1.
    \end{algorithmic}
\end{algorithm}

\pagebreak

\section{Household size distributions}\label{appendix: hh size dist}
We generated synthetic data for two very different household distributions: the UK Labour Force Survey (LFS) from 2023 and an unrealistic ``split'' distribution in which most households are either composed of 2 or the maximum (6) number of individuals. In the former case, we took the UK LFS household size distribution, removed the probability of selecting a household of size 1 and re-normalised the distribution, obtaining the probability $P_{\text{UK}}^k$ that a randomly selected household in our population is of size $k$ ($k=2,\dots, 6$). We then derived the size-weighted distribution, i.e.\ we made the probability of a household included dataset having $k$ contacts proportional to $kP_{UK}^k$. This is the household distribution we would expect to see in our study if an infection spreading between household could be described well by a homogeneously mixing model in the general population, so that each individual in the population is equally likely be infected, and further assuming the same probability for each case to be detected and have their household included in the study. The split distribution was selected so that the mean of the size-weighted distribution was the same as the size-weighted distribution based on the UK LFS. See Figure \ref{fig: hh size distributions} for the size-weighted distributions used.
\pagebreak

\section{Procedure for selecting appropriate papers from \textcite{madewell_household_2022}}
From the full list of papers that estimated secondary attack rate in \textcite{madewell_household_2022}, we first filtered out papers, so that the remaining ones were compatible with the model assumptions and that we could extract a low-information dataset, according to the following criteria:
\begin{enumerate}
    \item In each household, all contacts must have been tested or screened for symptoms;
    \item There was a procedure for identifying the primary and co-primary cases, e.g.\ based on symptom onset or epidemiological investigation;
    \item If co-primary cases were suspected, those individuals were not removed without the rest of the household being removed, as this would have left the household partially depleted;
    \item There was sufficient information reported to deduce a total number of contacts, cases and households.
\end{enumerate}

Papers that received ad-hoc consideration were the following:
\begin{itemize}
    \item \textcite{hsu_household_2021}: SAR was reported for family clusters, not households. Here, the work was included, with clusters treated as households.
    \item \textcite{singanayagam_community_2022}: SARs were reported for 127 households and 138 index cases. It was not possible to disentangle the contacts in households with a single primary cases from the information provided and so this paper was not included.
    \item \textcite{dub_high_2022}: Different SARs, based on PCR testing and symptoms, were reported. We included the data from PCR testing.
    \item \textcite{tanaka_sars-cov-2_2022}: The number of households with a single primary case in the study was not clear and so this paper was not included.
    \item \textcite{park_coronavirus_2020}: The number of households was not explicitly stated and so was estimated from the mean household size and total number of contacts as $225/2.3\approx98$ households. The paper was included.
\end{itemize}

\pagebreak

%% file: Sections/supplement.tex
\begin{figure}[h!]
    \centering
    \includegraphics[width=\linewidth]{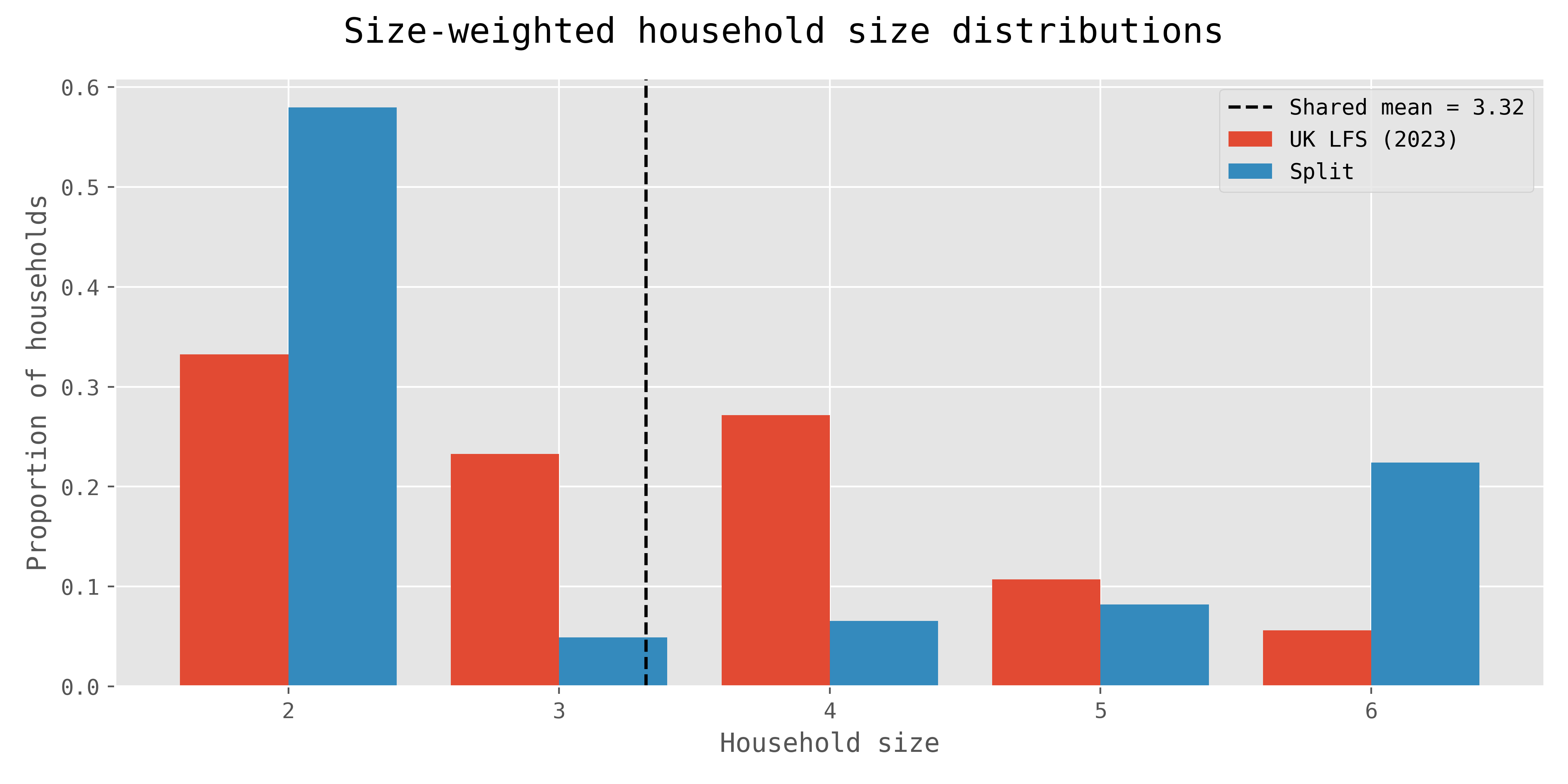}
    \caption{Size-weighted household size distributions that were used to generate synthetic data. The distribution for the UK LFS (2023) \cite{ons_families_2024} and the ``split'' distribution are shown in red and blue, respectively. Both distributions share a mean final size, after size weighting, of 3.32 which is shown by the vertical black dotted line. For the purpose of this paper households reported to have 6+ individuals in the UK LFS are assumed to have size exactly 6.}
    \label{fig: hh size distributions}
\end{figure}

\pagebreak
\begin{figure}
    \centering
    \includegraphics[width=\linewidth]{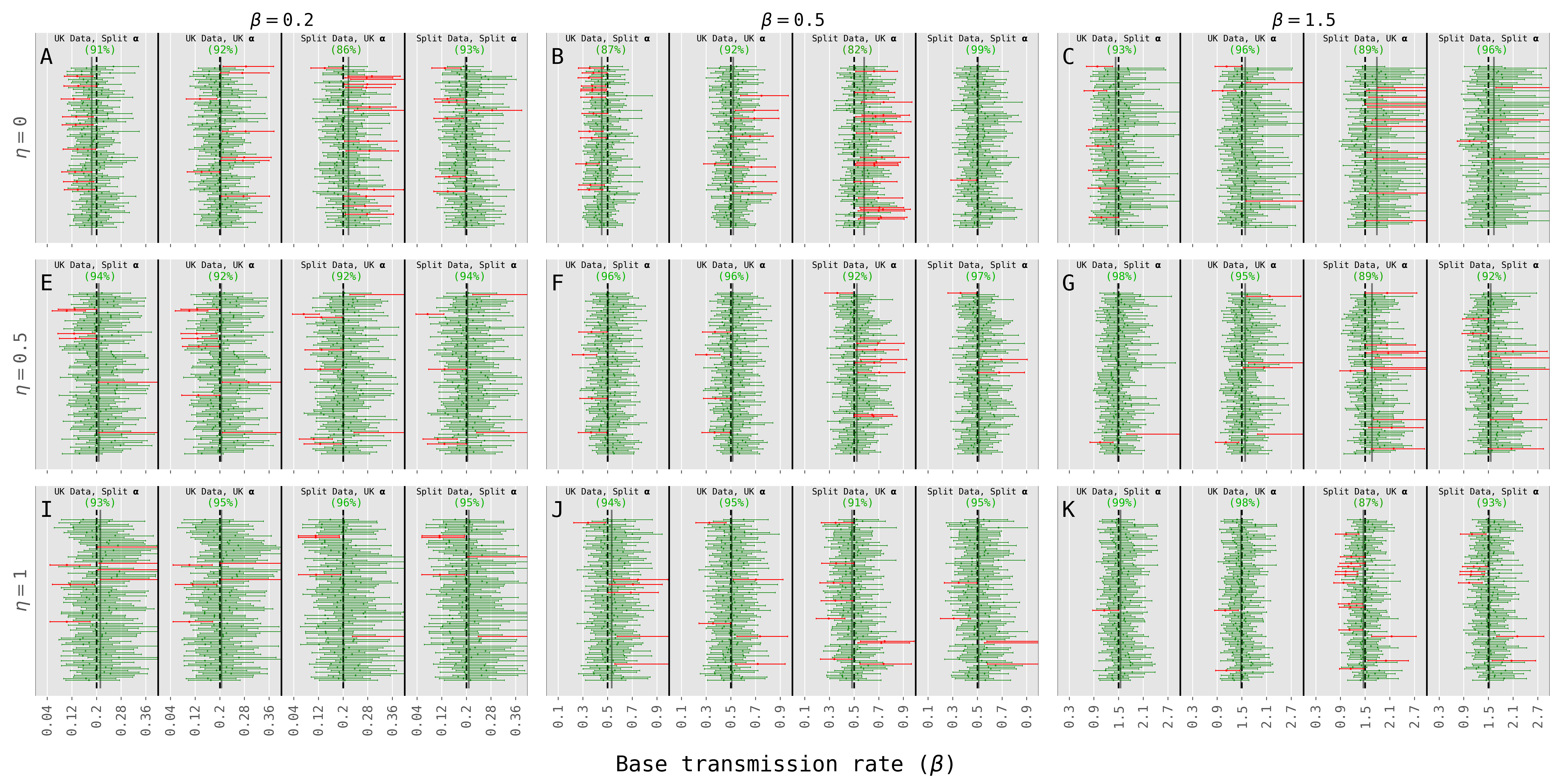}
    \caption{In each subplot, 95\% credible intervals for the posteriors of $\beta$ are shown for 100 low information synthetic datasets per household size distribution. Each dataset is generated using the Ball model ($I\sim \text{Gamma}(2,2)$) and $N=100$ household sizes sampled either from the UK LFS (2023) or the ``split'' household size distribution (see Appendix \ref{appendix: hh size dist} for details), and $\beta$ is re-estimated separately using $\boldsymbol{\alpha}$ proportional to each of the UK LFS and split household size distributions ($\alpha_0 = 100$). Each subplot row shows results for synthetic data generated with different base transmission rate ($\beta$) and each column shows results for different density mixing parameter ($\eta$). The value of $\beta$ used to generate the data is shown by the vertical dotted line in each plot and credible intervals are plotted in green if this value is contained in them and red otherwise. The average mean estimate of $\beta$ across all 100 datasets is shown by the grey vertical line. The percentage of synthetic datasets for which the real value is contained within the 95\% credible interval is shown above each set of credible intervals. All fits were done with $I\sim \text{Gamma}(2,2)$ for a known $\eta$ and so only $\beta$ was inferred.}
    \label{fig:synthetic data validation 100}
\end{figure}

\begin{figure}
    \centering
    \includegraphics[width=\linewidth]{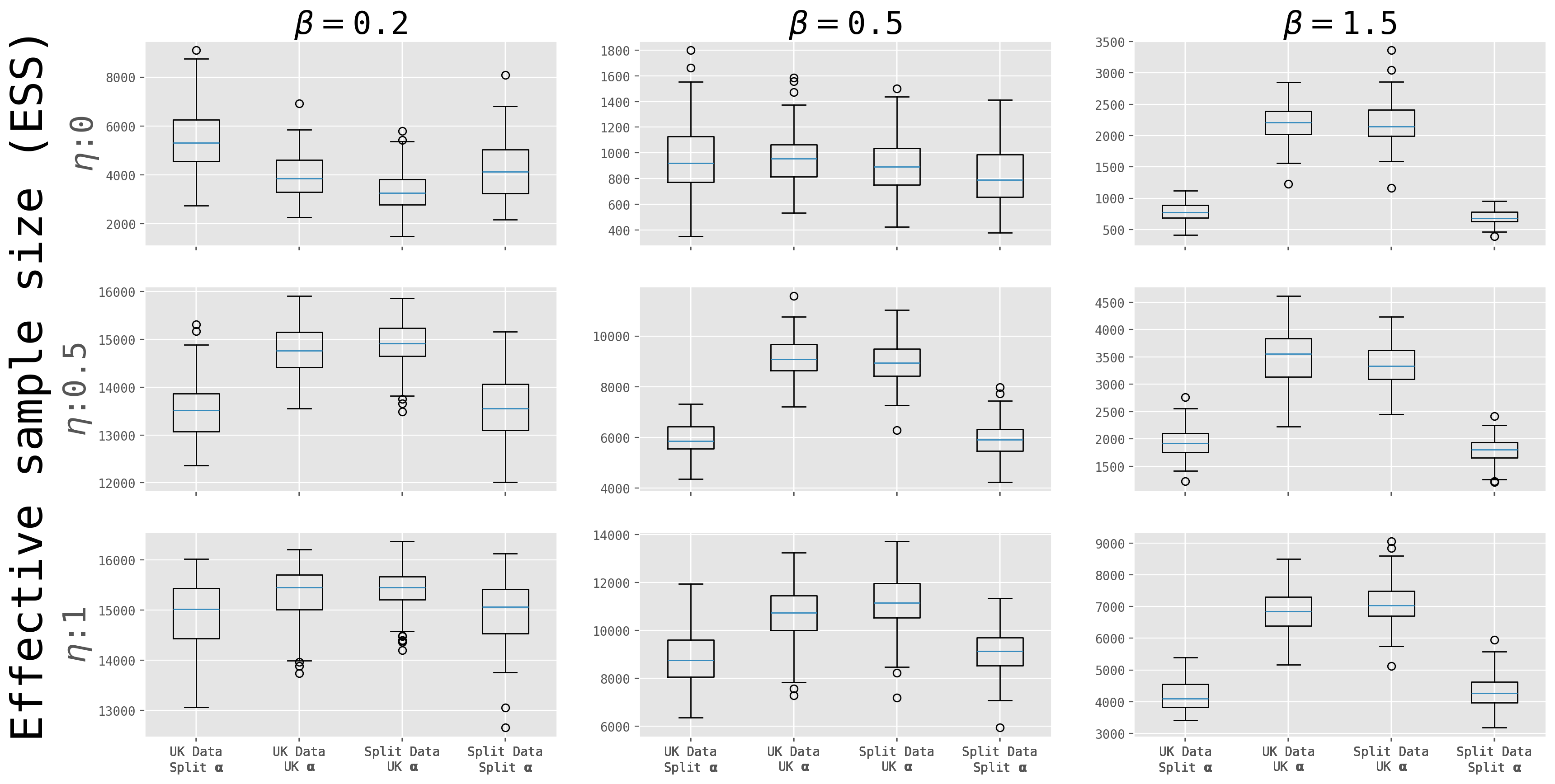}
    \caption{Box and whisker plots showing the distribution of ESS values for different synthetic datasets with $N=1000$.}
    \label{fig:ESS box and whisker 1000}
\end{figure}
\begin{figure}
    \centering
    \includegraphics[width=\linewidth]{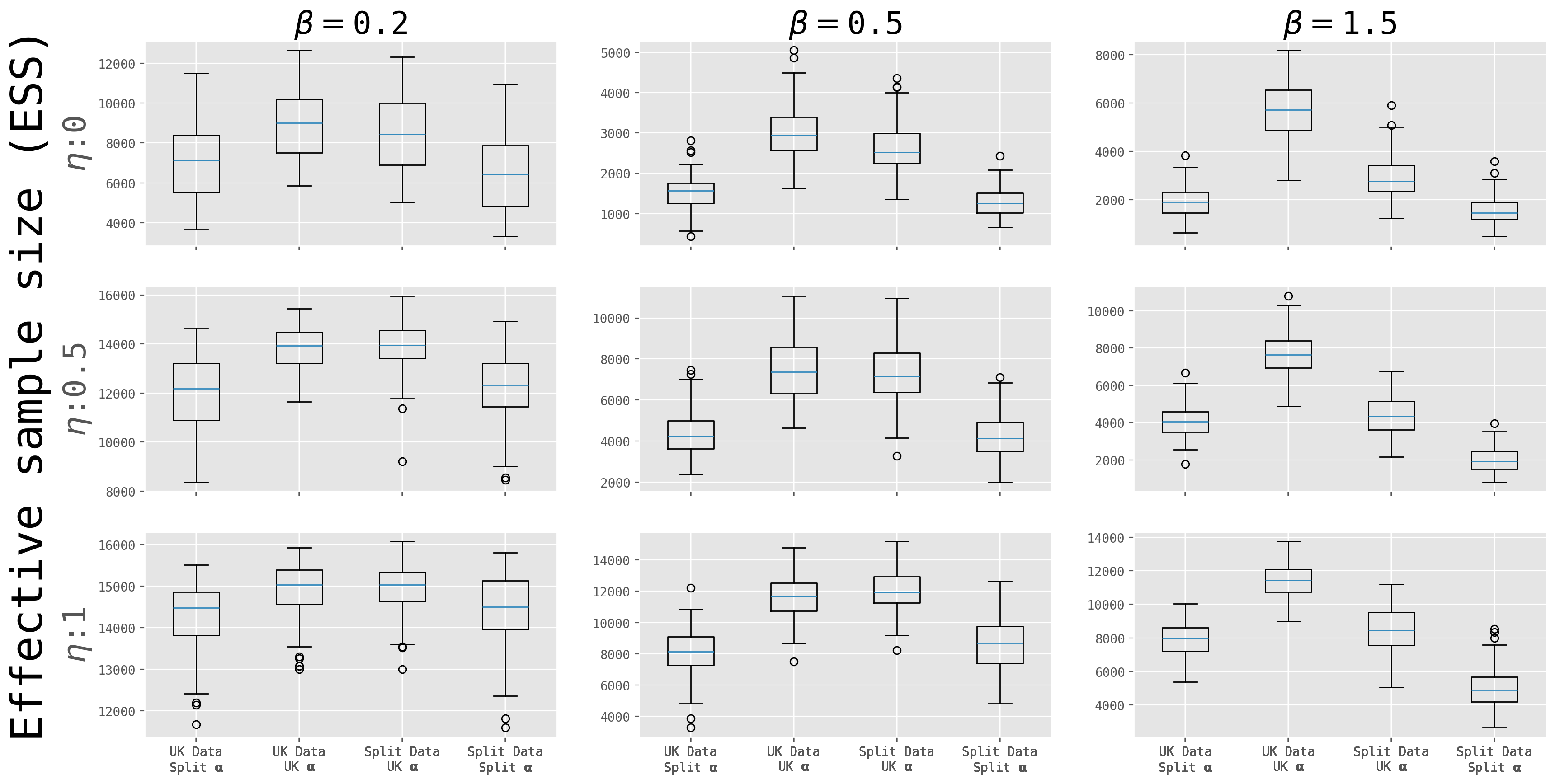}
    \caption{Box and whisker plots showing the distribution of ESS values for different synthetic datasets with $N=100$.}
    \label{fig:ESS box and whisker 100}
\end{figure}

\pagebreak
\begin{figure}[p]
    \centering
    \includegraphics[width=\linewidth]{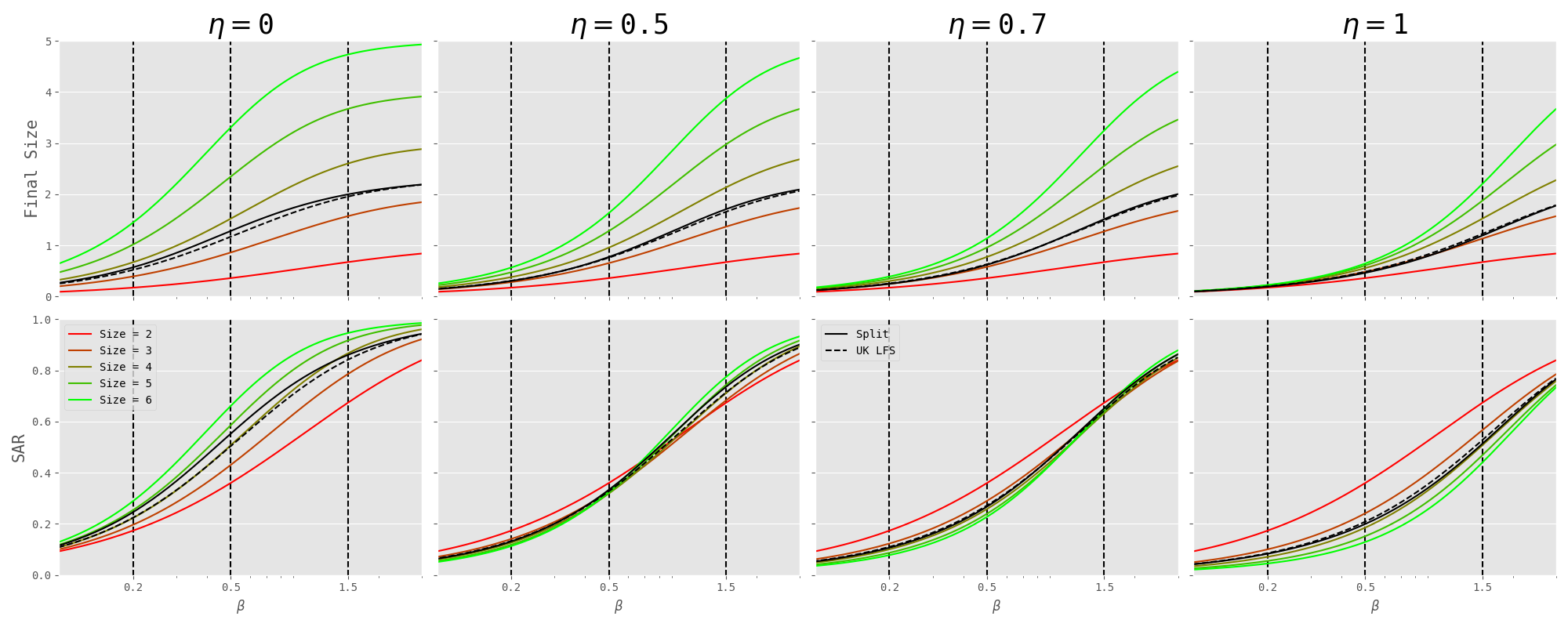}
    \caption{In this figure we compare the mean final size (top row) and mean SAR (bottom row) for households of different sizes (2--6) for different transmission rates ($\beta$ -- shown on the x-axis on a log-scale). Vertical dashed lines are shown at the values of $\beta$ considered in Figure \ref{fig: synthetic data validation 1000} synthetic data validation. Each column shows results for a different value of $\eta$, ranging from $\eta = 0$ (density dependent mixing), $\eta = 0.5$, $\eta=0.7$ and $\eta=1$ (frequency dependent mixing). We also show the mean final size/SAR for the two different household size distributions(UK LFS and ``split'') Figure \ref{fig: hh size distributions}.}
    \label{fig:FS/SAR supplement}
\end{figure}
\pagebreak

\begin{figure}[p]
    \centering
    \includegraphics[width=\linewidth]{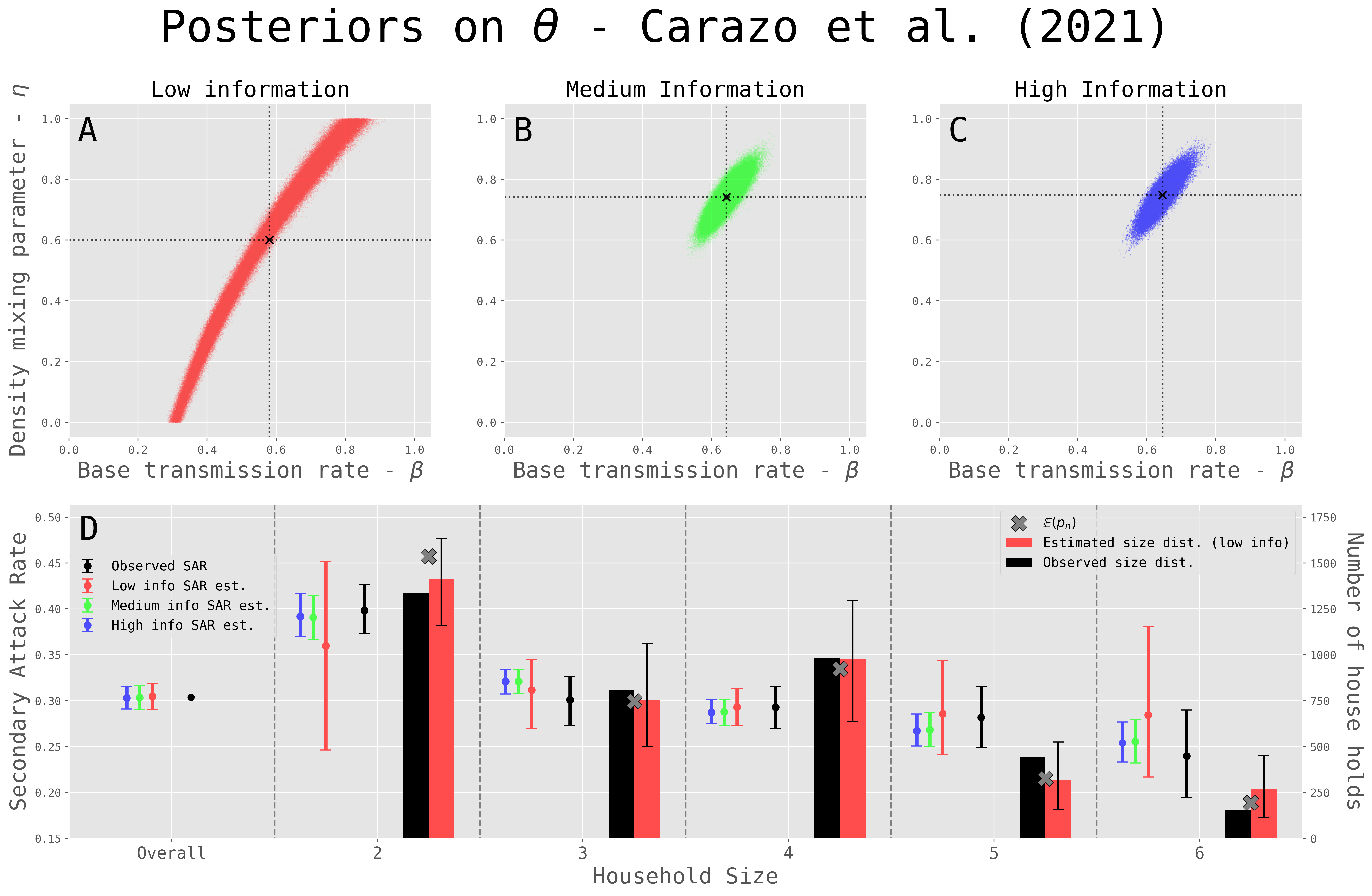}
    \caption{Panels A, B, and C show the posterior distributions of $\theta$ obtained by fitting the low- (red), medium- (green), and high-information (blue) versions of the data from \textcite{carazo_characterization_2021}, respectively. Panel D displays the secondary attack rate (SAR) implied by each posterior alongside the observed SAR (black) for each household size and overall; error bars represent 95\% credible intervals, with those for the observed SAR estimated by bootstrapping. Bars indicate the number of households of each size in the low-information fit (red) compared with the observed distribution (black). Grey Xs indicate the expected number of households of each size implied by the choice of $\boldsymbol{\alpha}$ The infectious period is assumed to be $I \sim \text{Exponential}(1)$.}
    \label{fig: carazo Scatter Markov}
\end{figure}

\begin{figure}[p]
    \centering
    \includegraphics[width=\linewidth]{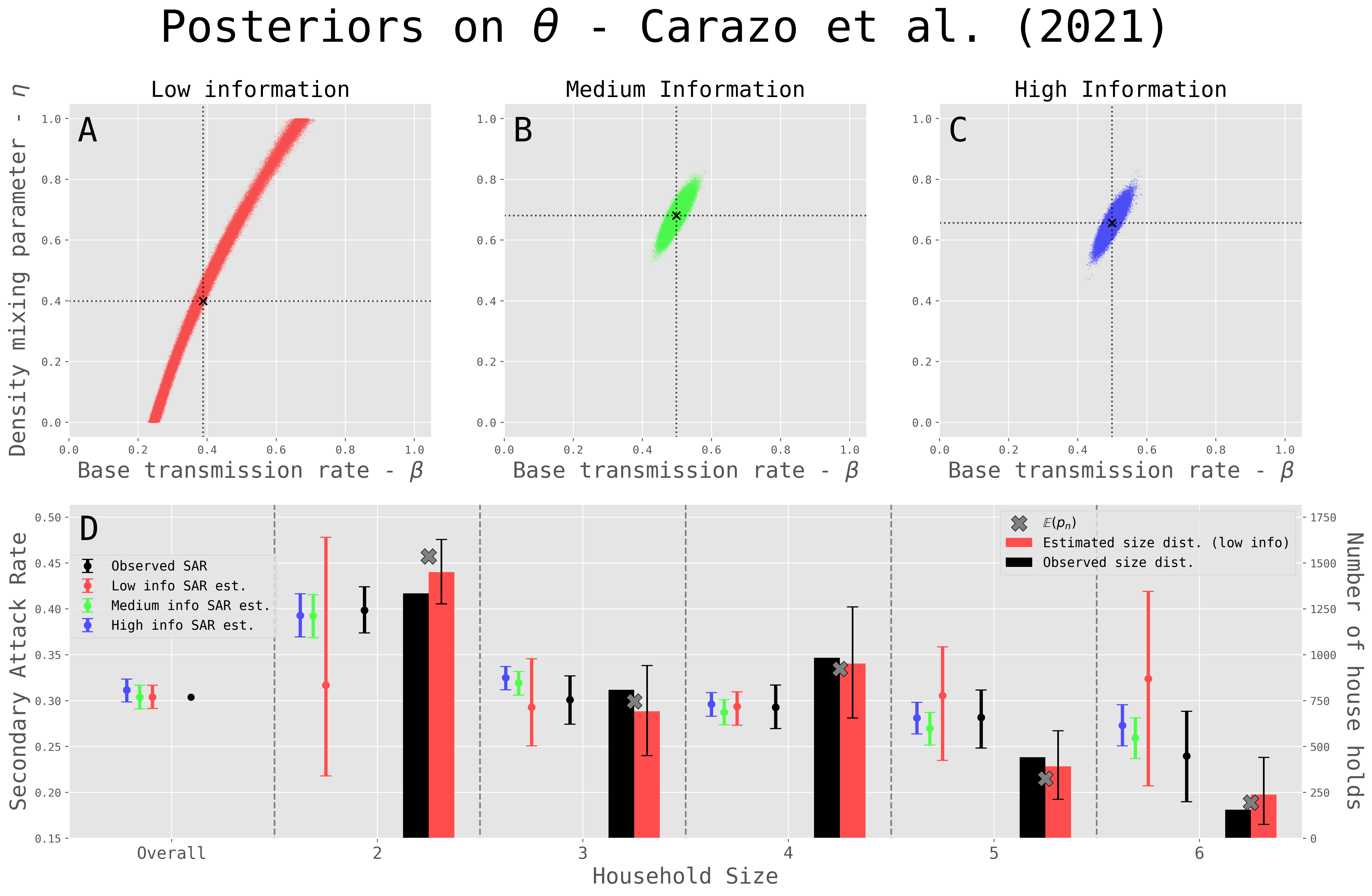}
    \caption{Panels A, B, and C show the posterior distributions of $\theta$ obtained by fitting the low- (red), medium- (green), and high-information (blue) versions of the data from \textcite{carazo_characterization_2021}, respectively. Panel D displays the secondary attack rate (SAR) implied by each posterior alongside the observed SAR (black) for each household size and overall; error bars represent 95\% credible intervals, with those for the observed SAR estimated by bootstrapping. Bars indicate the number of households of each size in the low-information fit (red) compared with the observed distribution (black). Grey Xs indicate the expected number of households of each size implied by the choice of $\boldsymbol{\alpha}$ The infectious period is assumed to be fixed to $I = 1$.}
    \label{fig: carazo Scatter Fixed}
\end{figure}

\pagebreak

\begin{figure}[p]
    \centering
    \includegraphics[width=\linewidth]{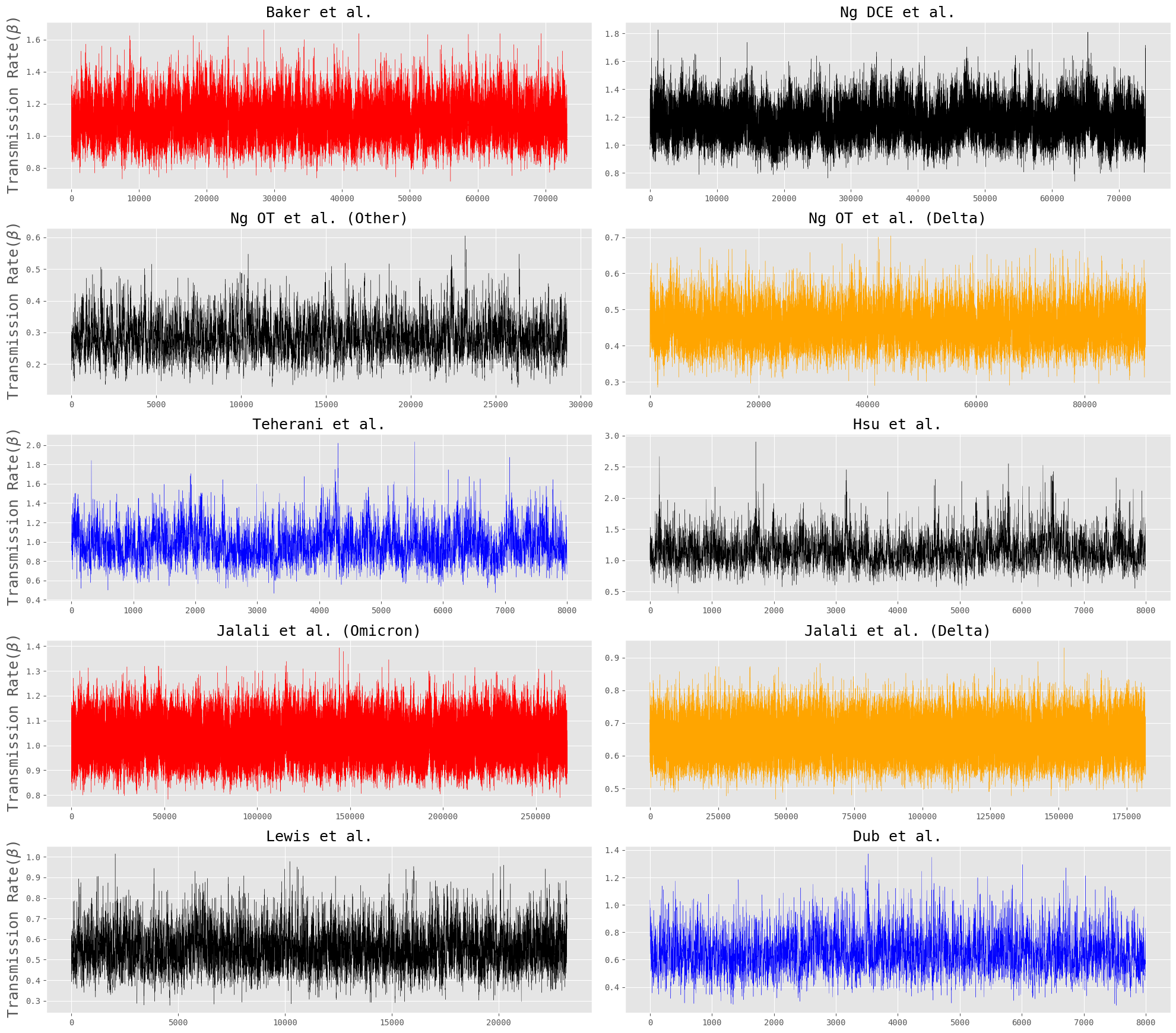}
    \caption{Trace plots for $\beta$ for each of the papers reporting low information datasets taken from Madewell \emph{et al.} (datasets 1-10). Each plot is colour-coded by the variant, as in Figure \ref{fig:MW_beta_fits}: blue for pre-Alpha, green for Alpha, orange for Delta, red for Omicron and black for Unspecified/Other. Note that the chain lengths ($x$-axis ranges) vary significantly between panels and that all chains mix well: those that appear not to mix well are just much shorter than the others and correspond to studies with small numbers of households.} 
    \label{fig:mw traceplot1}
\end{figure}
\begin{figure}[p]
    \centering
    \includegraphics[width=\linewidth]{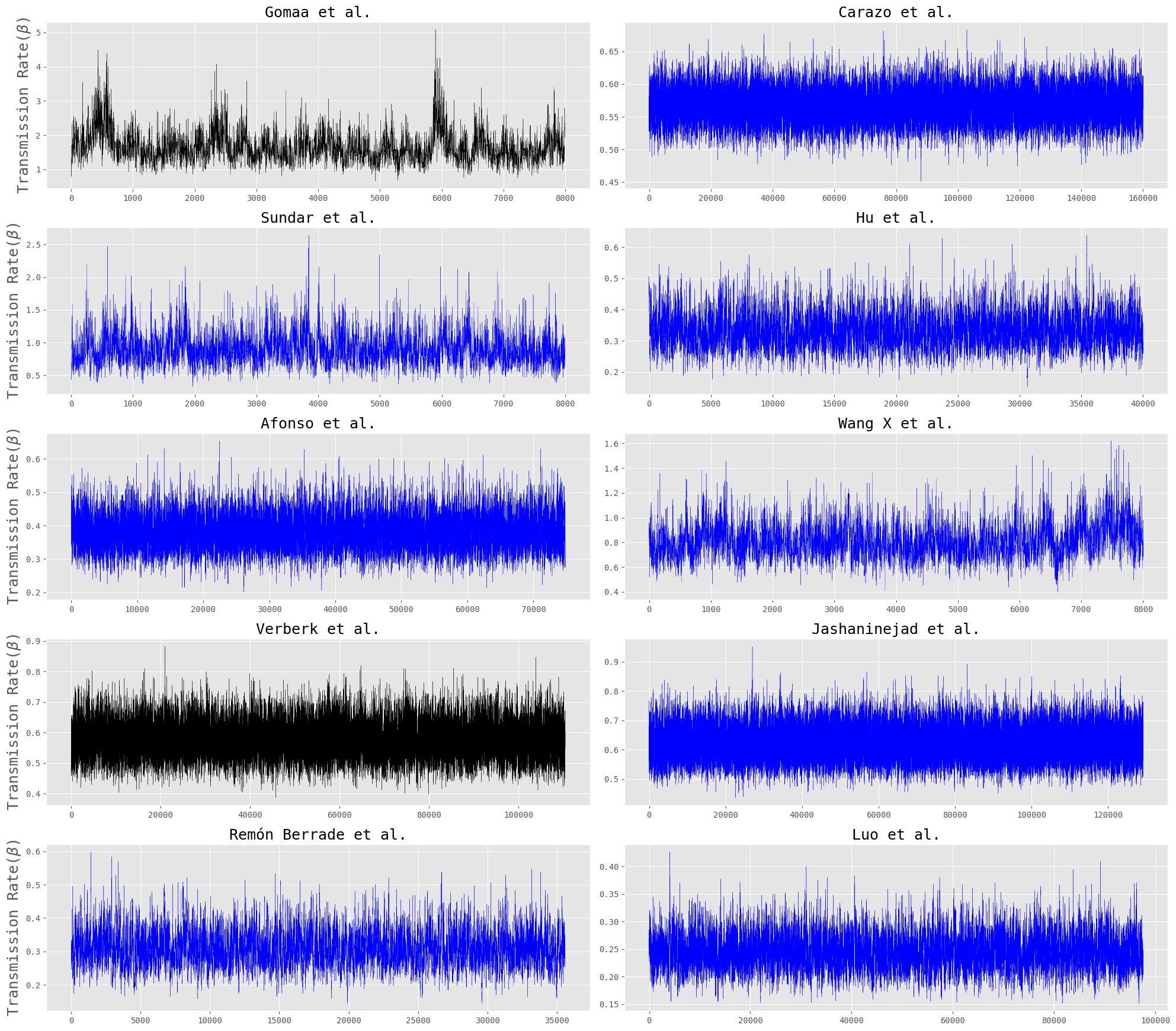}
    \caption{Trace plots for $\beta$ for each of the papers reporting low information datasets taken from Madewell \emph{et al.} (datasets 11-20). Each plot is colour-coded by the variant, as in Figure \ref{fig:MW_beta_fits}: blue for pre-Alpha, green for Alpha, orange for Delta, red for Omicron and black for Unspecified/Other. Note that the chain lengths ($x$-axis ranges) vary significantly between panels and that all chains mix well: those that appear not to mix well are just much shorter than the others and correspond to studies with small numbers of households.}
    \label{fig:mw traceplot2}
\end{figure}

\begin{figure}[p]
    \centering
    \includegraphics[width=\linewidth]{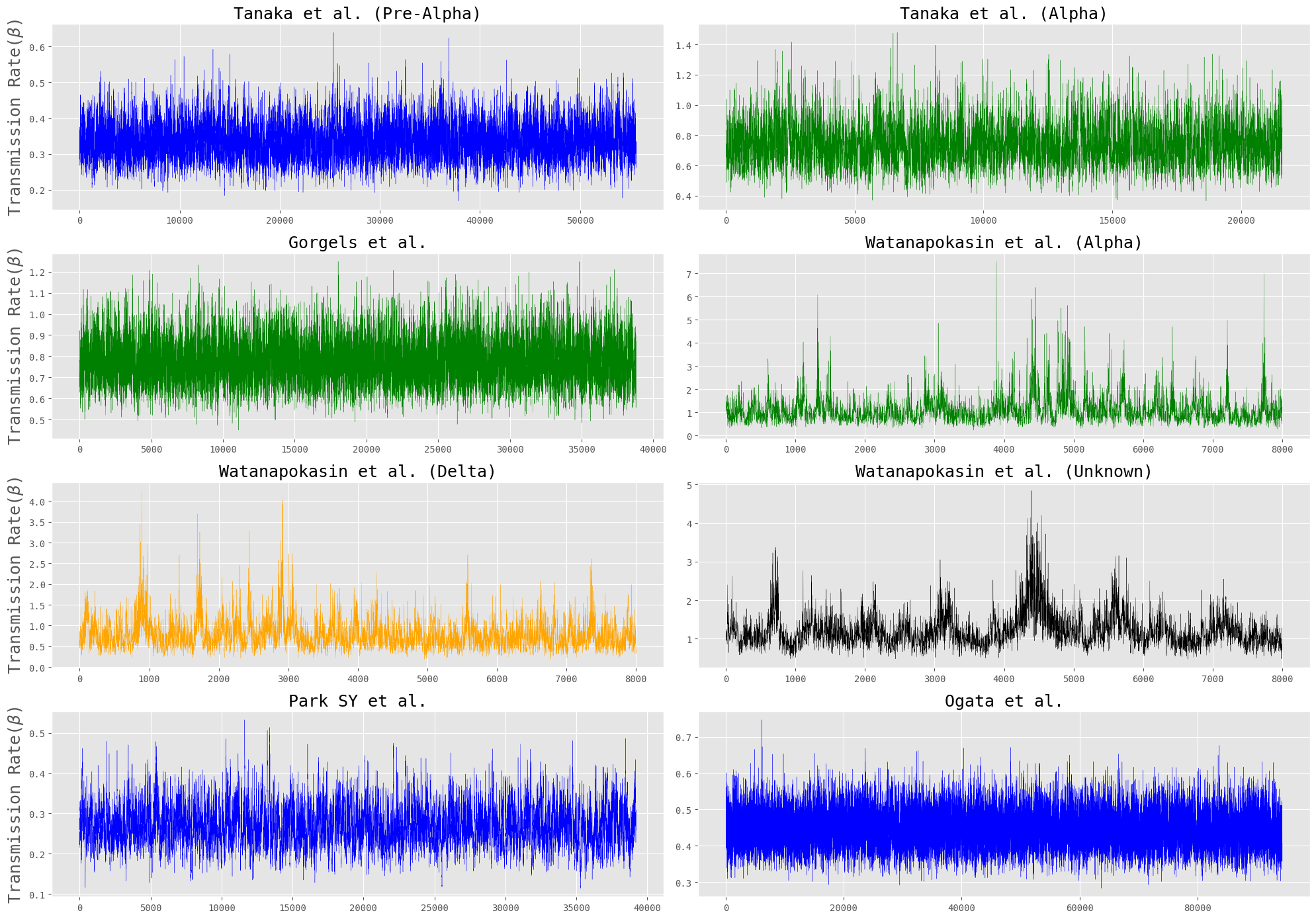}
    \caption{Trace plots for $\beta$ for each of the papers reporting low information datasets taken from Madewell \emph{et al.} (datasets 21-28). Each plot is colour-coded by the variant, as in Figure \ref{fig:MW_beta_fits}: blue for pre-Alpha, green for Alpha, orange for Delta, red for Omicron and black for Unspecified/Other. Note that the chain lengths ($x$-axis ranges) vary significantly between panels and that all chains mix well: those that appear not to mix well are just much shorter than the others and correspond to studies with small numbers of households.}
    \label{fig:mw traceplot3}
\end{figure}